\pdfoutput=1
\documentclass[11pt,letterpaper]{article}

\usepackage[margin=1in]{geometry}
\usepackage[utf8]{inputenc}
\usepackage[T1]{fontenc}
\usepackage{lmodern}
\usepackage{microtype}
\usepackage{setspace}
\usepackage{amsmath,amssymb,amsfonts,amsthm,mathtools}
\usepackage{bm}
\usepackage{graphicx}
\usepackage{array,booktabs,longtable}
\usepackage{tikz}
\usetikzlibrary{arrows.meta,fit,positioning}
\usepackage{xcolor}
\usepackage[authoryear,round]{natbib}
\usepackage[
  colorlinks=true,
  linkcolor=blue!60!black,
  citecolor=blue!60!black,
  urlcolor=blue!60!black,
  pdftitle={Distributional Graphical Models},
  pdfauthor={Jintao Wen, Yang Ni, and Weixuan Zhu}
]{hyperref}

\numberwithin{equation}{section}
\allowdisplaybreaks

\newtheorem{proposition}{Proposition}
\theoremstyle{definition}

\newtheorem{remark}{Remark}

\newcommand{\dgm}{DGM}

\DeclareMathOperator{\logit}{logit}

\definecolor{diagblue}{RGB}{218,232,252}
\definecolor{withinwarm}{RGB}{255,242,204}
\definecolor{betweengreen}{RGB}{213,232,212}

\title{\Large\bfseries Distributional Graphical Models}

\author{%
  Jintao Wen$^{1}$, \quad
  Yang Ni$^{3,*}$, \quad
  and Weixuan Zhu$^{1,2,*}$\\[0.8em]
  \footnotesize\itshape $^{1}$Department of Statistics and Data Science, School of Economics, Xiamen University, Xiamen, China\\
  \footnotesize\itshape $^{2}$Wang Yanan Institute for Studies in Economics, Xiamen University, Xiamen, China\\
  \footnotesize\itshape $^{3}$Department of Statistics and Data Sciences, The University of Texas at Austin, Austin, TX, USA\\[0.45em]
  \footnotesize\normalfont Email: \texttt{wenjintao1@gmail.com}\\
  \footnotesize\normalfont $^{*}$Corresponding authors: \texttt{yang.ni@austin.utexas.edu}, \texttt{zhuweixuan@gmail.com}
}

\date{}

\begin{document}

\maketitle

\begin{abstract}
Multivariate distributional data have become increasingly popular; yet statistical methods for studying their conditional independence have not been well researched. Current graphical-model methodology for distribution-valued nodes is restricted to univariate distributions and separates distribution estimation from graph inference. We develop a Bayesian distributional graphical model for learning the conditional-independence graph among general distribution-valued variables. At each node, subject-specific distributions are represented by a finite mixture model. The mixture component parameters are shared across subjects, whereas the mixing weights vary by subject. We transform the mixing weights into latent Euclidean coordinates using logistic stick breaking. We then model the stacked coordinates with a Gaussian graphical model and use a block spike-and-slab prior for edge selection. The hierarchy propagates uncertainty in the distributional representations into graph inference. We establish identifiability of the shared-mixture representation and show that zero between-node precision blocks correspond to conditional independence between the associated distribution-valued variables. Posterior inference uses a partially collapsed Metropolis-within-Gibbs sampler with P\'olya--Gamma augmentation and column-wise precision-matrix updates. Simulations demonstrate improved graph recovery over baseline methods and robustness to mixture-kernel misspecification. An application to the OneK1K single-cell RNA sequencing data identifies a conditional-dependence graph among immune cell types from donor-specific multivariate expression distributions.

\medskip
\noindent\textbf{Keywords:} Distributional data; Finite mixture model; Gaussian graphical model; Single-cell RNA sequencing; Spike-and-slab prior.
\end{abstract}

\section{Introduction}
\label{sec:introduction}

Advances in high-throughput assays, digital sensing, and large-scale data collection increasingly yield many observations within each study unit. When scientific interest concerns unit-to-unit variation, a probability distribution naturally arises as a unit-level summary. This is known as the distributional data, in which each unit-specific distribution is observed through a finite sample \citep{PetersenMuller2016,PetersenZhangKokoszka2022, ZhouMuller2024}. Examples include intraday return distributions across trading days, physical-activity distributions across individuals, and gene-expression distributions across cells within subjects. Reducing these samples to a mean or mean vector can obscure dispersion, asymmetry, and multimodality. For example, in single-cell studies, aggregation can also mask subpopulation structure and discard cell-resolved information relevant to cell--cell communication \citep{KorthauerEtAl2016,WilkEtAl2024}. We therefore should regard the unit-specific distributions as the observations of interest.

Associations among distribution-valued variables have been studied through distribution-to-distribution regression, which relates a response distribution to one or more predictor distributions. Existing approaches include nonparametric regression and methods based on optimal transport, Wasserstein geometry, and the Fisher--Rao metric \citep{OlivaPoczosSchneider2013,GhodratiPanaretos2022,ChenLinMuller2023,ZhaoDattaTangZipunnikovCaffo2023}. Recent work extends this setting to multivariate distributions and multiple multivariate predictors \citep{NguyenNiMueller2025,NguyenNiMueller2026}. These methods characterize associations through specified predictor--response relationships.

Our focus is instead on modeling conditional independence among several distribution-valued variables. Such structure is naturally represented by an undirected graphical model, where a missing edge indicates that two nodes are conditionally independent given the remaining nodes. Existing graphical-model methodology has focused predominantly on finite-dimensional Euclidean random vectors, particularly Gaussian graphical models \citep{Dempster1972,Lauritzen1996, FriedmanHastieTibshirani2008,Wang2015SSSL}. Related work extends graphical modeling to Gaussian random functions and more general functional data \citep{ZhuStrawnDunson2016,QiaoGuoJames2019,SoleaLi2022}. Functional graphical models treat each node-level observation as a random function over an index domain; in our setting, the observation at each node is instead a probability distribution observed through a finite sample.

Very few graphical models have been developed for non-Euclidean random objects such as probability distributions. To our knowledge, the only published method is that of \citet{WuWangLu2025DistributionGraph}. Their procedure consists of three separate steps. First, each unit-specific univariate density is estimated from its observed sample using Gaussian kernel smoothing. Second, the M-LDQ embedding \citep{ZhaoWangLu2022MLDQ} represents the estimated density by its median and log quantile-density function, corresponding to position and shape, respectively; the shape function is approximated by a finite basis expansion. Third, correlation matrices are constructed separately for the position and shape representations, and partial-correlation tests within the PC-stable algorithm yield two graphs whose union defines the distributional graph. The method is restricted to univariate distributions. Its sequential construction treats the estimated densities and finite-dimensional feature representations as fixed inputs to graph estimation, and hence uncertainty from the first two stages is not propagated into graph inference. Moreover, the final graph combines separate position and shape analyses rather than arising from a joint probability model for the distribution-valued variables.

Motivated by these specific limitations and the general lack of distributional graphical models, we propose a Bayesian distributional graphical model (\dgm{}) for multivariate distributions. At each node, a subject-specific distribution is represented by a finite mixture model with mixture component shared across subjects but mixing weights specific to each subject, which capture the between-subject distributional variation. For demonstration, we choose Student-$t$ as the mixture component, which accommodates heavy tails and can reduce the need for additional components used solely to model tail behavior \citep{PeelMcLachlan2000}. We establish that the shared-component representation is identifiable up to a node-wise common permutation.

We use logistic stick breaking to map each subject-specific mixing-weight vector bijectively to a latent Euclidean coordinate block, and the stacked blocks are modeled jointly by a Gaussian graphical model. A zero between-node block is equivalent to conditional independence between the corresponding latent-coordinate blocks given the remaining blocks. Mixture identifiability and the bijectivity of the stick-breaking map extend this equivalence to conditional independence between the distribution-valued variables given the remaining variables. We extend the graphical spike-and-slab prior \citep{Wang2015SSSL} to block selection and adapt its column-at-a-time Gibbs update for the latent precision matrix. P\'olya--Gamma augmentation \citep{PolsonScottWindle2013,LindermanJohnsonAdams2015} yields conditionally Gaussian updates for the stick-breaking coordinates. The resulting hierarchical model jointly estimates the subject-specific distributions and the conditional-independence graph. We illustrate the effectiveness of \dgm{} with simulations and apply \dgm{} to the OneK1K single-cell eQTL dataset \citep{YazarEtAl2022} to infer a conditional-dependence graph among immune cell types from their donor-specific multivariate expression distributions.

The remainder of the article is organized as follows. Section~\ref{sec:model} formulates the hierarchical distributional graphical model and its prior. Section~\ref{sec:inference} describes posterior computation and graph estimation. Section~\ref{sec:simulations} presents simulation studies comparing the proposed method with mean-based and multistage alternatives. Section~\ref{sec:application} applies \dgm{} to population-scale single-cell transcriptomic data. Section~\ref{sec:discussion} concludes with a discussion. Proofs, additional posterior-computation details, simulation settings and sensitivity results, and application details are provided in Appendices~\ref{app:posterior-computation}--\ref{app:application-details}.

\section{Model}
\label{sec:model}

We consider data from \(N\) subjects, each providing a sample at each of \(p\) nodes. At each node, \dgm{} represents each subject-specific distribution using mixture components shared across subjects and subject-specific mixing weights, and links the transformed weights through a latent Gaussian graphical model with block sparsity. The resulting graph represents conditional dependence among the nodes, as captured by their subject-specific distributions, conditional on the remaining nodes.

\subsection{Mixture Representation}
\label{sec:mixture}

For node \(j\in\{1,\ldots,p\}\), the observations from subject \(i\) form the sample
\[
    \mathcal X_i^{(j)}
    =
    \left\{
        \mathbf x_{i,c}^{(j)}
    \right\}_{c=1}^{N_i^{(j)}}
    \subset \mathbb R^{d_j},
\]
where \(c\) indexes observations within the sample. The observation dimension \(d_j\) may vary across nodes, whereas the sample size \(N_i^{(j)}\) may vary across both subjects and nodes.

For \(K\geq2\), let
\(
    \Delta^{K-1}
    =
    \left\{
        \bm\pi\in[0,1]^K:
        \sum_{k=1}^K\pi_k=1
    \right\}
\)
denote the simplex, and let \(\operatorname{int}(\Delta^{K-1})\) denote its interior, in which every coordinate is positive. We model the distribution underlying each sample as a finite mixture of multivariate Student-\(t\) components. Student-\(t\) kernels accommodate heavy tails and reduce sensitivity to atypical observations \citep{PeelMcLachlan2000}. They can accommodate heavy-tailed behavior with fewer components than Gaussian mixtures.

Let \(K_j\) denote the number of components at node \(j\), and let
\(
    \bm\Theta^{(j)}
    =
    \left\{
        (\bm\mu_{j,k},\bm\Sigma_{j,k})
    \right\}_{k=1}^{K_j}
\)
denote the component parameters shared across subjects. For subject \(i\), let
\[
    \bm\pi_i^{(j)}
    =
    \left(
        \pi_{i,1}^{(j)},\ldots,\pi_{i,K_j}^{(j)}
    \right)^\top
    \in
    \operatorname{int}
    \left(
        \Delta^{K_j-1}
    \right)
\]
denote the subject-specific mixing weights. Let \(t_{\nu_t,d_j}(\mathbf x\mid\bm\mu,\mathbf S)\) denote the \(d_j\)-dimensional Student-\(t\) density with \(\nu_t\) degrees of freedom, location \(\bm\mu\), and scale matrix \(\mathbf S\). Conditional on \(\bm\pi_i^{(j)}\) and \(\bm\Theta^{(j)}\), the observations in \(\mathcal X_i^{(j)}\) are independent draws from
\begin{equation}
\label{eq:mix_t_density}
    p\left(
        \cdot
        \mid
        \bm\pi_i^{(j)},\bm\Theta^{(j)}
    \right)
    =
    \sum_{k=1}^{K_j}
    \pi_{i,k}^{(j)}
    \,
    t_{\nu_t,d_j}
    \left(
        \cdot
        \mid
        \bm\mu_{j,k},
        c_{\nu_t}\bm\Sigma_{j,k}
    \right),
\end{equation}
where \(c_{\nu_t} = \frac{\nu_t-2}{\nu_t}\) for \(\nu_t>2\) ensures that \(\bm\Sigma_{j,k}\) is the marginal covariance matrix of component \(k\).

For ease of posterior computation, we augment model \eqref{eq:mix_t_density} with component indicators $z_{i,c}^{(j)}$ and latent scales $\xi_{i,c}^{(j)}$:
\begin{align*}
    z_{i,c}^{(j)}
    \mid
    \bm\pi_i^{(j)}
    &\sim
    \operatorname{Categorical}
    \left(\bm\pi_i^{(j)}\right),\\
    \xi_{i,c}^{(j)}
    &\sim
    \operatorname{Gamma}
    \left(
        \frac{\nu_t}{2},
        \frac{\nu_t}{2}
    \right),\\
    \mathbf x_{i,c}^{(j)}
    \mid
    z_{i,c}^{(j)}=k,\xi_{i,c}^{(j)},
    \bm\mu_{j,k},\bm\Sigma_{j,k}
    &\sim
    \mathcal N_{d_j}
    \left(
        \bm\mu_{j,k},
        \frac{c_{\nu_t}\bm\Sigma_{j,k}}
             {\xi_{i,c}^{(j)}}
    \right).
\end{align*}
Marginalizing out $z_{i,c}^{(j)}$ and $\xi_{i,c}^{(j)}$ recovers \eqref{eq:mix_t_density}.

Sharing component parameters across subjects defines a common node-specific dictionary, so that between-subject distributional variation is represented by the mixing weights on a common basis. This structure permits information sharing when within-subject sample sizes are small or unbalanced. Allowing component locations and covariances to vary freely across subjects would confound changes in the mixture basis with changes in the mixing weights, complicating graph inference and interpretation. Since our shared-parameter mixture models are less standard, we provide the identifiability property below.

\begin{proposition}[Identifiability up to a node-wise common permutation]
\label{prop:mixture_identifiability}
Fix a node \(j\), the number of components \(K_j\), and \(\nu_t>2\). Let \(f_i^{(j)}\) and \(\widetilde f_i^{(j)}\), \(i=1,\ldots,N\), be two collections of densities of the form~\eqref{eq:mix_t_density}, based respectively on the shared dictionaries
\[
    \bm\Theta^{(j)}
    =
    \left\{
        (\bm\mu_{j,k},\bm\Sigma_{j,k})
    \right\}_{k=1}^{K_j}
    \qquad\text{and}\qquad
    \widetilde{\bm\Theta}^{(j)}
    =
    \left\{
        (\widetilde{\bm\mu}_{j,k},
         \widetilde{\bm\Sigma}_{j,k})
    \right\}_{k=1}^{K_j},
\]
and on subject-specific mixing weights \(\bm\pi_i^{(j)}\) and \(\widetilde{\bm\pi}_i^{(j)}\).

Assume that all covariance matrices are positive definite, the component parameter pairs are pairwise distinct within each dictionary, and
\[
    \bm\pi_i^{(j)},
    \widetilde{\bm\pi}_i^{(j)}
    \in
    \operatorname{int}
    \left(
        \Delta^{K_j-1}
    \right).
\]
If
\[
    f_i^{(j)}
    =
    \widetilde f_i^{(j)}
    \quad\text{almost everywhere}
\]
for every \(i=1,\ldots,N\), then there exists a permutation \(\sigma_j\) of $\{1,\dots,K_j\}$, common to all subjects at node \(j\), such that
\[
    \left(
        \widetilde{\bm\mu}_{j,\sigma_j(k)},
        \widetilde{\bm\Sigma}_{j,\sigma_j(k)}
    \right)
    =
    \left(
        \bm\mu_{j,k},
        \bm\Sigma_{j,k}
    \right)
    \qquad\text{and}\qquad
    \widetilde\pi_{i,\sigma_j(k)}^{(j)}
    =
    \pi_{i,k}^{(j)}
\]
for every \(i=1,\ldots,N\) and \(k=1,\ldots,K_j\).
\end{proposition}

The proof is provided in Appendix~\ref{app:proof-proposition-1}.

\subsection{Distributional Graph via Latent Coordinates}
\label{sec:latent_graph}

We characterize conditional dependence among the \(p\) nodes through the joint variation of their subject-specific mixing weights across subjects. Because these weights lie on probability simplices, we map them to Euclidean coordinates using logistic stick breaking \citep{LindermanJohnsonAdams2015}. Let \(\sigma(v)=\{1+\exp(-v)\}^{-1}\) and \(r_j=K_j-1\). For \(\mathbf V_i^{(j)} =(V_{i1}^{(j)},\ldots,V_{i,r_j}^{(j)})^\top\), define
\[
\pi_{i,k}^{(j)} = \sigma\left(V_{ik}^{(j)}\right) \prod_{\ell=1}^{k-1} \left\{1-\sigma\left(V_{i\ell}^{(j)}\right)\right\}, \qquad k=1,\ldots,K_j-1,
\]
and
\[
\pi_{i,K_j}^{(j)} = \prod_{\ell=1}^{K_j-1} \left\{1-\sigma\left(V_{i\ell}^{(j)}\right)\right\}.
\]
We write this bijection as $\bm\pi_i^{(j)} = \operatorname{SB}_{K_j}\left(\mathbf V_i^{(j)}\right)$. Its sequential-binomial representation enables the P\'olya--Gamma augmentation used in Section~\ref{sec:inference}.

Let $Q=\sum_{j=1}^p r_j$, and concatenate the subject-specific latent coordinates as
\[
    \mathbf V_i
    =
    \left[
        \left(\mathbf V_i^{(1)}\right)^\top,
        \ldots,
        \left(\mathbf V_i^{(p)}\right)^\top
    \right]^\top
    \in\mathbb R^Q .
\]
We model their joint variation across subjects by a Gaussian graphical model,
\[
    \mathbf V_i
    \mid
    \bm\mu_V,\bm\Omega
    \sim
    \mathcal N_Q
    \left(
        \bm\mu_V,
        \bm\Omega^{-1}
    \right),
    \qquad
    i=1,\ldots,N,
\]
where \(\bm\Omega\) is a positive definite precision matrix. Let \(\mathcal I_j\subset\{1,\ldots,Q\}\) denote the coordinate indices for node \(j\), and let \(\mathcal I_{-j}=\{1,\ldots,Q\}\setminus\mathcal I_j\), so that \(\mathbf V_i^{(j)}=\mathbf V_{i,\mathcal I_j}\). For distinct nodes \(a\) and \(b\), the between-node precision block is \(\bm\Omega_{\mathcal I_a,\mathcal I_b} \in \mathbb R^{r_a\times r_b}\).

\begin{remark}[Conditional independence at the distribution level]
\label{rem:distributional_graph}
For a fixed subject \(i\), let \(F_i^{(j)}\) denote the distribution at node \(j\), and let \(R_{ab}:=\{1,\ldots,p\}\setminus\{a,b\}\). Once component labels are fixed, Proposition~\ref{prop:mixture_identifiability} and the bijectivity of the logistic stick-breaking transformation imply a one-to-one correspondence between \(F_i^{(j)}\) and \((\bm\Theta^{(j)},\mathbf V_i^{(j)})\). Because the component-parameter collections \(\{\bm\Theta^{(j)}\}_{j=1}^p\) are independent across nodes and independent of the latent coordinates, under the model-implied law conditional on \((\bm\mu_V,\bm\Omega)\),
\[
\begin{aligned}
\bm\Omega_{\mathcal I_a,\mathcal I_b}=\mathbf 0
\quad
&\Longleftrightarrow
\quad
\mathbf V_i^{(a)}
\perp\!\!\!\perp
\mathbf V_i^{(b)}
\mid
\{\mathbf V_i^{(j)}:j\in R_{ab}\}
\\
&\Longleftrightarrow
\quad
F_i^{(a)}
\perp\!\!\!\perp
F_i^{(b)}
\mid
\{F_i^{(j)}:j\in R_{ab}\}.
\end{aligned}
\]
The first equivalence follows from the Gaussian block-Markov property. For the second, the independence of \(\bm\Theta^{(1)},\ldots,\bm\Theta^{(p)}\) across nodes and from \(\mathbf V_i\) implies that conditional independence among the latent-coordinate blocks is equivalent to conditional independence among the corresponding pairs \((\bm\Theta^{(j)},\mathbf V_i^{(j)})\). The node-wise one-to-one correspondence between these pairs and \(F_i^{(j)}\) then gives the conditional-independence relation at the distribution level.
\end{remark}

\subsection{Prior Specification}
\label{sec:priors}

\paragraph{Mixture component priors.}
We assign Normal--Inverse-Wishart priors to the shared Student-\(t\) component parameters:
\[
\bm\Sigma_{j,k} \sim \mathcal{IW}_{d_j}(\nu_j,\bm\Psi_j), \qquad \bm\mu_{j,k}\mid\bm\Sigma_{j,k} \sim \mathcal N_{d_j} \left( \bm m_{0j}, \bm\Sigma_{j,k}/\kappa_j \right).
\]
We use an empirical Bayes approach, where $\bm m_{0j}$ and $\bm\Psi_j$ are specified as the pooled sample mean and the pooled sample covariance at node $j$, respectively.

To mitigate the label switching issue of mixture models, we restrict the means of the mixture components within each node according to the first coordinate:
\[
\mu_{j,1,1} < \mu_{j,2,1} < \cdots < \mu_{j,K_j,1},
\]
where \(\mu_{j,k,1}\) is the first coordinate of \(\bm\mu_{j,k}\).

\paragraph{Latent graph mean.}
We assign the latent graph mean a conjugate Gaussian prior conditional on the precision matrix,
\[
    \bm\mu_V
    \mid
    \bm\Omega
    \sim
    \mathcal N_Q
    \left(
        \bm m_V,
        (\kappa_V\bm\Omega)^{-1}
    \right),
    \qquad
    \kappa_V>0.
\]

\paragraph{Block graph prior.}
We specify the graph-layer prior jointly over three disjoint sets of precision-matrix entries: the diagonal entries, the within-node off-diagonal entries, and the between-node entries. The diagonal entries determine the conditional precision of each latent stick-breaking coordinate. The within-node off-diagonal entries characterize conditional dependence among the latent coordinates underlying the mixing weights at the same node. The between-node entries characterize conditional dependence between the latent-coordinate vectors underlying the mixing weights of two nodes, given the remaining nodes. All prior factors are restricted jointly to the positive-definite cone \(\mathcal S_Q^+\).

\begin{figure}[htbp]
\centering
\resizebox{0.62\linewidth}{!}{%
\begin{tikzpicture}[x=0.55cm,y=0.55cm,font=\scriptsize]

\fill[betweengreen] (0,0) rectangle (6,-6);

\foreach \k in {0,2,4}{
    \fill[withinwarm]
        (\k,-\k) rectangle ++(2,-2);
}

\foreach \k in {0,...,5}{
    \fill[diagblue]
        (\k,-\k) rectangle ++(1,-1);
}

\draw[step=1,gray!70,thin] (0,-6) grid (6,0);

\foreach \k in {0,2,4,6}{
    \draw[line width=0.8pt] (\k,0) -- (\k,-6);
    \draw[line width=0.8pt] (0,-\k) -- (6,-\k);
}

\node at (0.5,-0.5) {\bfseries D};
\node at (1.5,-1.5) {\bfseries D};
\node at (2.5,-2.5) {\bfseries D};
\node at (3.5,-3.5) {\bfseries D};
\node at (4.5,-4.5) {\bfseries D};
\node at (5.5,-5.5) {\bfseries D};

\node at (1.5,-0.5) {\bfseries W};
\node at (0.5,-1.5) {\bfseries W};
\node at (3.5,-2.5) {\bfseries W};
\node at (2.5,-3.5) {\bfseries W};
\node at (5.5,-4.5) {\bfseries W};
\node at (4.5,-5.5) {\bfseries W};

\node[fill=betweengreen,inner sep=1pt] at (3,-1) {\bfseries B};
\node[fill=betweengreen,inner sep=1pt] at (5,-1) {\bfseries B};
\node[fill=betweengreen,inner sep=1pt] at (1,-3) {\bfseries B};
\node[fill=betweengreen,inner sep=1pt] at (5,-3) {\bfseries B};
\node[fill=betweengreen,inner sep=1pt] at (1,-5) {\bfseries B};
\node[fill=betweengreen,inner sep=1pt] at (3,-5) {\bfseries B};

\node at (1,0.42) {node 1};
\node at (3,0.42) {node 2};
\node at (5,0.42) {node 3};

\node[rotate=90] at (-0.42,-1) {node 1};
\node[rotate=90] at (-0.42,-3) {node 2};
\node[rotate=90] at (-0.42,-5) {node 3};

\draw[fill=diagblue]
    (7.0,-0.25) rectangle (7.65,-0.90);
\node[anchor=west] at (7.85,-0.575)
    {\(\mathrm D:\ q_{\mathrm{diag}}\)};

\draw[fill=withinwarm]
    (7.0,-1.35) rectangle (7.65,-2.00);
\node[anchor=west] at (7.85,-1.675)
    {\(\mathrm W:\ q_{\mathrm{within}}\)};

\draw[fill=betweengreen]
    (7.0,-2.45) rectangle (7.65,-3.10);
\node[anchor=west] at (7.85,-2.775)
    {\(\mathrm B:\ q_{\mathrm{between}}\)};

\end{tikzpicture}%
}
\caption{Schematic partition of the precision matrix for three nodes with \(r_1=r_2=r_3=2\). Entries labeled D determine coordinate-specific conditional precisions. Entries labeled W characterize within-node conditional dependence among the latent coordinates. Blocks labeled B characterize conditional dependence between the latent-coordinate vectors of two nodes. The factors \(q_{\mathrm{diag}}\), \(q_{\mathrm{within}}\), and \(q_{\mathrm{between}}\) govern these three sets, respectively.}
\label{fig:precision_prior_partition}
\end{figure}
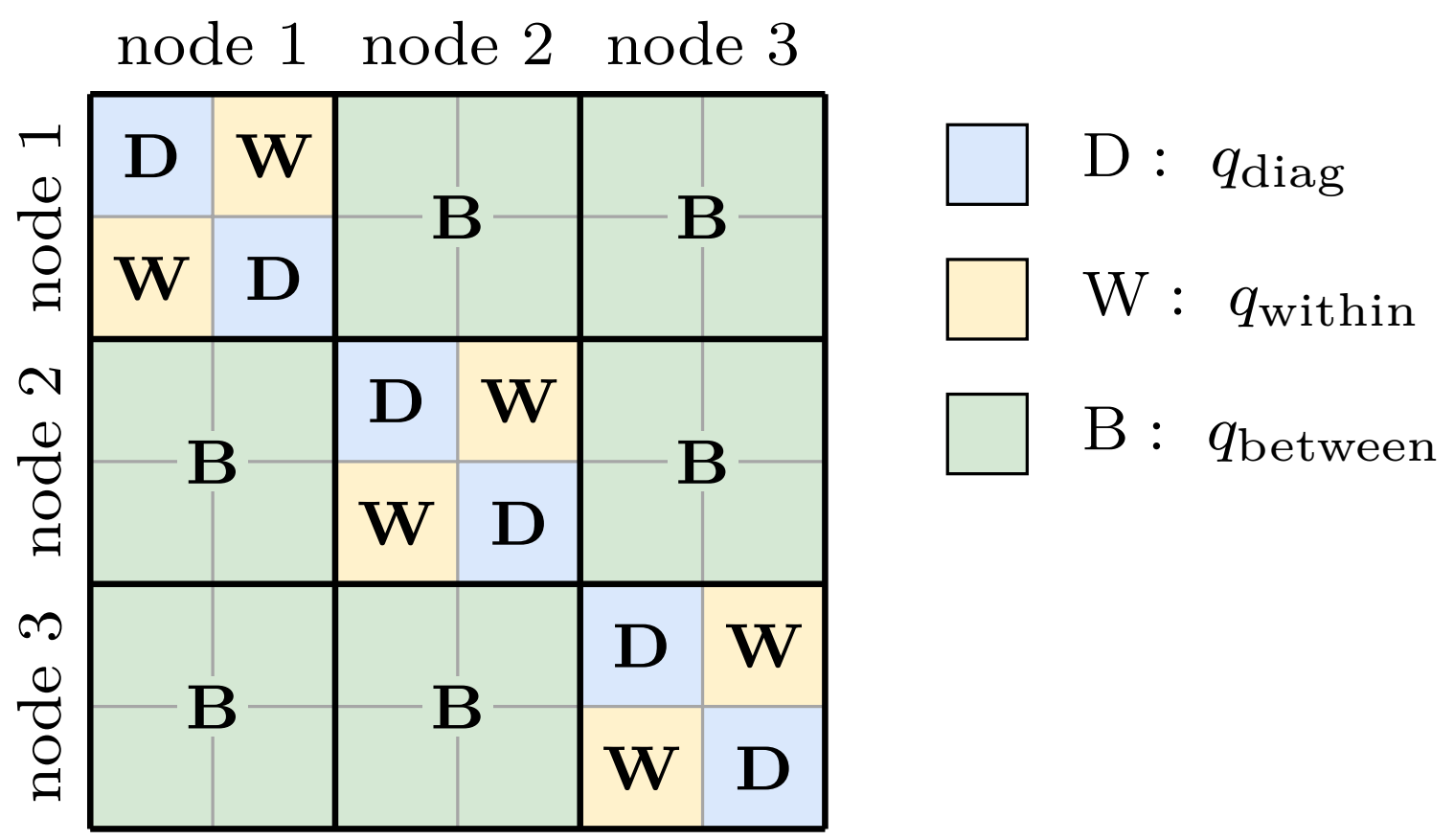

We summarize the graph-layer prior as
\[
p(\bm\Omega\mid\boldsymbol\vartheta)
\propto
q_{\mathrm{diag}}(\bm\Omega\mid\boldsymbol\vartheta)\,
q_{\mathrm{within}}(\bm\Omega\mid\boldsymbol\vartheta)\,
q_{\mathrm{between}}(\bm\Omega\mid\boldsymbol\vartheta)\,
\mathbf 1\{\bm\Omega\in\mathcal S_Q^+\},
\]
where \(\boldsymbol\vartheta\) collects the hyperparameters specified below. Figure~\ref{fig:precision_prior_partition} illustrates the three classes of entries and their corresponding prior factors.

For the diagonal entries, we define the prior factor
\[
    q_{\mathrm{diag}}(\bm\Omega\mid\boldsymbol\vartheta)
    =
    \prod_{s=1}^{Q}
    \exp
    \left(
        -\frac{\lambda_{\mathrm{diag}}}{2}\Omega_{ss}
    \right).
\]

For the within-node off-diagonal entries, we define
\[
    q_{\mathrm{within}}(\bm\Omega\mid\boldsymbol\vartheta)
    =
    \prod_{j=1}^{p}
    \prod_{\substack{s<t\\s,t\in\mathcal I_j}}
    \phi
    \left(
        \Omega_{st};
        0,\sigma_{\mathrm{within}}^2
    \right),
\]
where \(\phi(\cdot;m,s^2)\) denotes the normal density with mean \(m\) and variance \(s^2\).

For the between-node factor, consider distinct nodes \(a\) and \(b\), with coordinate indices \(u=1,\ldots,r_a\) and \(v=1,\ldots,r_b\). Let
\[
    \omega_{ab,uv}
    :=
    \left[
        \bm\Omega_{\mathcal I_a,\mathcal I_b}
    \right]_{uv}
\]
denote the \((u,v)\) entry of the corresponding between-node precision block.

We introduce a block multiplier $\alpha_{ab}$ and an entry-specific variable $\Lambda_{ab,uv}$, and write
\[
    \omega_{ab,uv}
    =
    \alpha_{ab}\Lambda_{ab,uv}
\]
to separate node-pair-level selection from within-block variation. Because the number of components \(K_j\) may vary across nodes, the dimensions of the corresponding between-node precision blocks may also differ. Applying shrinkage and selection directly to individual block entries would therefore make the representation of a node-level edge depend on the number of entries in the corresponding block. The shared multiplier \(\alpha_{ab}\) instead provides a common block-level scale that governs edge selection, while \(\Lambda_{ab,uv}\) accommodates entry-specific signs and relative magnitudes within the block.

Let \(G_{ab}=G_{ba}\in\{0,1\}\), with \(G_{aa}=0\), denote the spike-or-slab indicator for node pair \((a,b)\). Conditional on \(G_{ab}=g\), we assign
\[
    \alpha_{ab}\mid G_{ab}=g
    \sim
    \mathcal N(0,\tau_g^2),
    \qquad
    0<\tau_0<\tau_1,
    \qquad
    g\in\{0,1\}.
\]
For each entry within the block, we introduce a sign indicator $H_{ab,uv}$ with equal prior probabilities,
\[
    \Pr(H_{ab,uv}=+1)
    =
    \Pr(H_{ab,uv}=-1)
    =
    \frac{1}{2},
    \qquad
    \Lambda_{ab,uv}\mid H_{ab,uv}
    \sim
    \mathcal N(H_{ab,uv},1).
\]
Equivalently, conditional on \(\alpha_{ab}\) and \(H_{ab,uv}\),
\[
    \omega_{ab,uv}
    \mid
    \alpha_{ab},H_{ab,uv}
    \sim
    \mathcal N
    \left(
        \alpha_{ab}H_{ab,uv},
        \alpha_{ab}^2
    \right).
\]

Let
\[
    \mathbf G
    =
    \{G_{ab}:1\leq a<b\leq p\},
    \qquad
    \bm\alpha
    =
    \{\alpha_{ab}:1\leq a<b\leq p\},
\]
and let \(\mathbf H\) collect all \(H_{ab,uv}\). For fixed \(\mathbf G\), the unconstrained joint prior kernel is
\[
\begin{aligned}
&q_{\mathbf G}
\left(
    \bm\Omega,\bm\alpha,\mathbf H
\right)
\\
&\quad =
\prod_{s=1}^{Q}
\exp
\left(
    -\frac{\lambda_{\mathrm{diag}}}{2}\Omega_{ss}
\right)
\prod_{j=1}^{p}
\prod_{\substack{s<t\\s,t\in\mathcal I_j}}
\phi
\left(
    \Omega_{st};
    0,\sigma_{\mathrm{within}}^2
\right)
\\
&\qquad\quad\times
\prod_{1\leq a<b\leq p}
\left[
    \phi
    \left(
        \alpha_{ab};
        0,\tau_{G_{ab}}^2
    \right)
    \prod_{u=1}^{r_a}
    \prod_{v=1}^{r_b}
    \left\{
        \frac{1}{2}
        \phi
        \left(
            \omega_{ab,uv};
            \alpha_{ab}H_{ab,uv},
            \alpha_{ab}^2
        \right)
    \right\}
\right].
\end{aligned}
\]
The first, second, and third products correspond to \(q_{\mathrm{diag}}\), \(q_{\mathrm{within}}\), and \(q_{\mathrm{between}}\), respectively.

Restriction to \(\mathcal S_Q^+\) induces a graph-dependent normalizing constant. Suppressing dependence on the fixed prior hyperparameters, define
\[
    \mathcal C(\mathbf G)
    =
    \sum_{\mathbf H}
    \int
    q_{\mathbf G}
    \left(
        \bm\Omega,\bm\alpha,\mathbf H
    \right)
    \mathbf 1
    \left\{
        \bm\Omega\in\mathcal S_Q^+
    \right\}
    d\bm\Omega\,d\bm\alpha.
\]
The normalized conditional prior is therefore
\[
    p
    (\bm\Omega,\bm\alpha,\mathbf H\mid\mathbf G)
    =
    \frac{
        q_{\mathbf G}
        (\bm\Omega,\bm\alpha,\mathbf H)
        \mathbf 1
        \{\bm\Omega\in\mathcal S_Q^+\}
    }{
        \mathcal C(\mathbf G)
    }.
\]

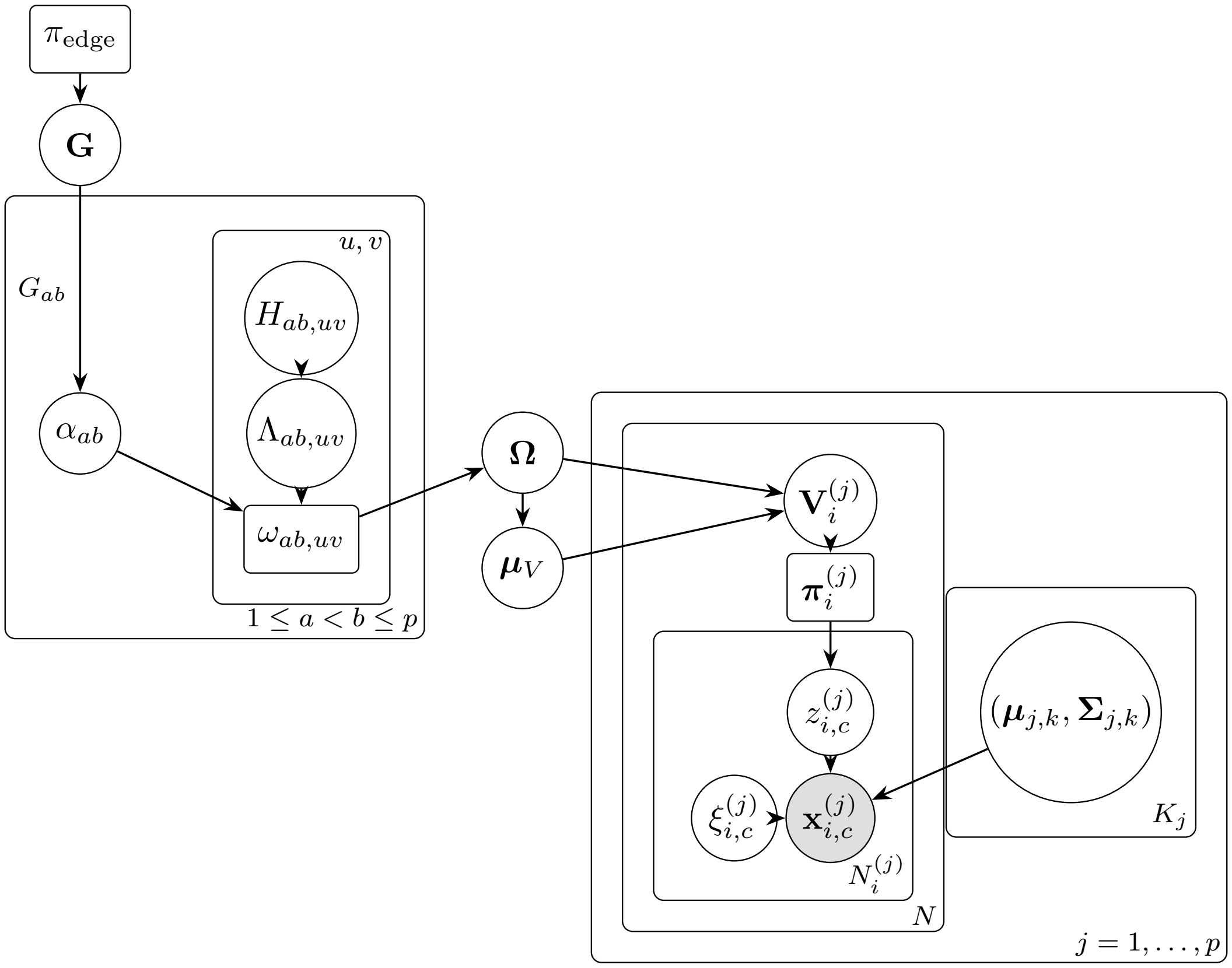
\begin{figure}[htbp]
\centering
\resizebox{0.84\linewidth}{!}{%
\begin{tikzpicture}[
>=Stealth,
latent/.style={
circle,
draw,
minimum size=24pt,
inner sep=1.5pt,
font=\small
},
obs/.style={
circle,
draw,
fill=gray!25,
minimum size=24pt,
inner sep=1.5pt,
font=\small
},
det/.style={
rectangle,
draw,
rounded corners=2pt,
minimum height=20pt,
inner xsep=4pt,
inner ysep=2pt,
font=\small
},
hyper/.style={
rectangle,
draw,
rounded corners=2pt,
minimum height=20pt,
inner xsep=4pt,
inner ysep=2pt,
font=\small
},
param/.style={
circle,
draw,
minimum size=24pt,
inner sep=1.5pt,
font=\small
},
plate/.style={
draw,
rounded corners=3pt
},
platelabel/.style={
font=\scriptsize,
inner sep=0pt
}
]

\node[hyper] (piE) at (-4.6,4.8) {$\pi_{\mathrm{edge}}$};
\node[latent] (Gvec) at (-4.6,3.7) {$\mathbf G$};
\node[latent] (alpha) at (-4.6,0.7) {$\alpha_{ab}$};
\node[latent] (Hab) at (-2.3,1.9) {$H_{ab,uv}$};
\node[latent] (Lam) at (-2.3,0.7) {$\Lambda_{ab,uv}$};
\node[det] (Oabuv) at (-2.3,-0.4) {$\omega_{ab,uv}$};

\draw[->,semithick] (piE) -- (Gvec);
\draw[->,semithick] (Gvec) -- node[left,font=\scriptsize] {$G_{ab}$} (alpha);
\draw[->,semithick] (Hab) -- (Lam);
\draw[->,semithick] (alpha) -- (Oabuv);
\draw[->,semithick] (Lam) -- (Oabuv);

\node[plate, fit=(Hab)(Lam)(Oabuv), inner sep=9pt] (plateUV) {};
\node[platelabel,anchor=north east] (uvIndex)
at ([xshift=-2pt,yshift=-2pt]plateUV.north east) {$u,v$};

\node[plate, fit=(alpha)(plateUV), inner sep=10pt] (plateAB) {};
\node[platelabel,anchor=south east] (abIndex)
at ([xshift=-2pt,yshift=2pt]plateAB.south east) {$1\leq a<b\leq p$};

\node[param] (Omega) at (0.0,0.5) {$\bm\Omega$};
\node[param] (muV) at (0.0,-0.7) {$\bm\mu_V$};

\draw[->,semithick] (Oabuv) -- (Omega);
\draw[->,semithick] (Omega) -- (muV);

\node[latent] (Vij) at (3.2,0.0) {$\mathbf V_i^{(j)}$};
\node[det] (pij) at (3.2,-0.9) {$\bm\pi_i^{(j)}$};
\node[latent] (zij) at (3.2,-2.2) {$z_{i,c}^{(j)}$};
\node[obs] (xij) at (3.2,-3.3) {$\mathbf x_{i,c}^{(j)}$};
\node[latent] (xi) at (2.2,-3.3) {$\xi_{i,c}^{(j)}$};
\node[param] (theta) at (5.7,-2.2) {$(\bm\mu_{j,k},\bm\Sigma_{j,k})$};

\draw[->,semithick] (Omega) -- (Vij);
\draw[->,semithick] (muV) -- (Vij);
\draw[->,semithick] (Vij) -- (pij);
\draw[->,semithick] (pij) -- (zij);
\draw[->,semithick] (zij) -- (xij);
\draw[->,semithick] (theta) -- (xij);
\draw[->,dashed,semithick] (xi) -- (xij);

\node[plate, fit=(zij)(xij)(xi), inner sep=11pt] (plateC) {};
\node[platelabel,anchor=south east] (obsIndex)
at ([xshift=-2pt,yshift=2pt]plateC.south east) {$N_i^{(j)}$};

\node[plate, fit=(Vij)(pij)(plateC), inner sep=9pt] (plateI) {};
\node[platelabel,anchor=south east] (subjectIndex)
at ([xshift=-2pt,yshift=2pt]plateI.south east) {$N$};

\node[plate, fit=(theta), inner sep=10pt] (plateK) {};
\node[platelabel,anchor=south east] (componentIndex)
at ([xshift=-2pt,yshift=2pt]plateK.south east) {$K_j$};

\node[plate, fit=(plateI)(plateK), inner sep=9pt] (plateJ) {};
\node[platelabel,anchor=south east] (nodeIndex)
at ([xshift=-2pt,yshift=2pt]plateJ.south east) {$j=1,\ldots,p$};

\end{tikzpicture}%
}
\caption{Hierarchical representation of \dgm{}.}
\label{fig:plate_diagram}
\end{figure}

For \(\pi_{\mathrm{edge}}\in(0,1)\), define the Bernoulli-product kernel
\[
    b_{\pi}(\mathbf G)
    =
    \prod_{1\leq a<b\leq p}
    \pi_{\mathrm{edge}}^{G_{ab}}
    (1-\pi_{\mathrm{edge}})^{1-G_{ab}},
    \qquad
    \mathcal C_{\pi}(\pi_{\mathrm{edge}})
    =
    \sum_{\mathbf G}
    \mathcal C(\mathbf G)b_{\pi}(\mathbf G).
\]
Following the normalizing-constant cancellation construction of \citet{Wang2015SSSL}, we specify
\begin{align*}
    p(\mathbf G\mid\pi_{\mathrm{edge}})
    &=
    \frac{
        \mathcal C(\mathbf G)b_{\pi}(\mathbf G)
    }{
        \mathcal C_{\pi}(\pi_{\mathrm{edge}})
    },\\
    p(\pi_{\mathrm{edge}})
    &\propto
    \mathcal C_{\pi}(\pi_{\mathrm{edge}})
    \pi_{\mathrm{edge}}^{a_\pi-1}
    (1-\pi_{\mathrm{edge}})^{b_\pi-1}.
\end{align*}
The second expression is a normalizing-constant-adjusted Beta hyperprior. Multiplying the preceding distributions cancels both \(\mathcal C(\mathbf G)\) and \(\mathcal C_{\pi}(\pi_{\mathrm{edge}})\), yielding the joint prior kernel
\[
p\left(
    \bm\Omega,\bm\alpha,\mathbf H,
    \mathbf G,\pi_{\mathrm{edge}}
\right)
\propto
q_{\mathbf G}(\bm\Omega,\bm\alpha,\mathbf H)\,
\mathbf 1\{\bm\Omega\in\mathcal S_Q^+\}\,
b_{\pi}(\mathbf G)\,
\pi_{\mathrm{edge}}^{a_\pi-1}
(1-\pi_{\mathrm{edge}})^{b_\pi-1}.
\]
The practical implication is that our posterior inference procedure does not need to evaluate these intractable normalizing constants.
The full hierarchical model is represented by the plate diagram in Figure~\ref{fig:plate_diagram}.

\section{Posterior Inference}
\label{sec:inference}

We use a partially collapsed Metropolis-within-Gibbs sampler. Each iteration first updates the shared Student-\(t\) component parameters, followed by the allocations and latent scales. It then updates the P\'olya--Gamma variables and latent coordinates, the columns of \(\bm\Omega\), the block-prior variables \((\mathbf H,\bm\alpha,\mathbf G)\), and \(\bm\mu_V\), in that order. The allocation update integrates out the Student-\(t\) scales, which are sampled immediately afterward. We use the induced Gaussian prior for the between-node precision entries directly, so \(\Lambda_{ab,uv}\) is not sampled separately. We also integrate out \(\pi_{\mathrm{edge}}\) when updating \(\mathbf G\). All blocks are updated by Gibbs or collapsed Gibbs sampling except the shared multipliers \(\alpha_{ab}\), which are updated by Metropolis--Hastings.

In this section, we present the DGM-specific transition kernels: the collapsed allocation--scale step, the P\'olya--Gamma latent-coordinate update, the Schur-complement update of $\bm\Omega$, and the collapsed edge-indicator update. Appendix~\ref{app:posterior-computation} provides the remaining conditional distributions, including the order-preserving component-parameter conditionals, the Gaussian full conditional for $\bm\mu_V$, the Bernoulli full conditional for \(H_{ab,uv}\), and the Metropolis--Hastings target for $\alpha_{ab}$. Hereafter, $\cdot$ denotes the observed data and the current values of all quantities retained in the corresponding conditional distribution.

\subsection{Mixture and Latent-Coordinate Updates}
\label{sec:posterior_computation}

Conditional on the current allocations and Student-\(t\) scales, we update each shared component by alternating full-conditional draws. The weighted sufficient statistics use observation weights \(\xi_{i,c}^{(j)}/c_{\nu_t}\). We first draw \(\bm\Sigma_{j,k}\) from its inverse-Wishart full conditional given the current \(\bm\mu_{j,k}\), and then draw \(\bm\mu_{j,k}\) from its Gaussian full conditional given the updated \(\bm\Sigma_{j,k}\). The first coordinate of each component mean is drawn from its truncated-normal conditional between the neighboring component means. Starting from an ordered configuration, these updates preserve the component ordering and thereby fix the component labels throughout the sampler.

Given the updated component parameters and current mixing weights, each allocation is sampled according to
\[
    \Pr\{z_{i,c}^{(j)}=k\mid\cdot\}
    \propto
    \pi_{i,k}^{(j)}
    t_{\nu_t,d_j}
    \left(
        \mathbf x_{i,c}^{(j)}
        \mid
        \bm\mu_{j,k},c_{\nu_t}\bm\Sigma_{j,k}
    \right),
    \qquad k=1,\ldots,K_j.
\]
This collapsed update integrates out $\xi_{i,c}^{(j)}$. The corresponding scale is then sampled immediately from
\[
    \xi_{i,c}^{(j)}
    \mid z_{i,c}^{(j)}=k,\cdot
    \sim
    \operatorname{Gamma}
    \left(
        \frac{\nu_t+d_j}{2},
        \frac{\nu_t+\delta_{i,c,k}^{(j)}/c_{\nu_t}}{2}
    \right),
\]
where
\[
    \delta_{i,c,k}^{(j)}
    =
    \left(\mathbf x_{i,c}^{(j)}-\bm\mu_{j,k}\right)^\top
    \bm\Sigma_{j,k}^{-1}
    \left(\mathbf x_{i,c}^{(j)}-\bm\mu_{j,k}\right).
\]

\paragraph{Stick-breaking latent-coordinate updates.}
Let $r_j=K_j-1$. For $k=1,\ldots,K_j$, define
\[
    n_{i,k}^{(j)}
    =
    \sum_{c=1}^{N_i^{(j)}}
    \mathbf 1\{z_{i,c}^{(j)}=k\}.
\]
For $\ell=1,\ldots,r_j$, define
\[
    R_{i,\ell}^{(j)}
    =
    \sum_{k=\ell}^{K_j}n_{i,k}^{(j)},
    \qquad
    \kappa_{i,\ell}^{(j)}
    =
    n_{i,\ell}^{(j)}-\frac{1}{2}R_{i,\ell}^{(j)}.
\]
P\'olya--Gamma augmentation \citep{PolsonScottWindle2013,LindermanJohnsonAdams2015} introduces
\[
    \zeta_{i,\ell}^{(j)}
    \mid R_{i,\ell}^{(j)},V_{i\ell}^{(j)}
    \sim
    \operatorname{PG}
    \left(R_{i,\ell}^{(j)},V_{i\ell}^{(j)}\right).
\]
Let
\[
    \bm\kappa_i^{(j)}
    =
    \left(
        \kappa_{i,1}^{(j)},\ldots,\kappa_{i,r_j}^{(j)}
    \right)^\top,
    \qquad
    \mathbf Z_i^{(j)}
    =
    \operatorname{diag}
    \left(
        \zeta_{i,1}^{(j)},\ldots,\zeta_{i,r_j}^{(j)}
    \right).
\]
Combining the augmented likelihood with the conditional Gaussian graphical model gives
\[
    \mathbf V_i^{(j)}\mid\cdot
    \sim
    \mathcal N_{r_j}
    \left(
        \mathbf A_{ij}^{-1}\bm\eta_{ij},
        \mathbf A_{ij}^{-1}
    \right),
\]
where
\begin{align*}
\mathbf A_{ij} &= \bm\Omega_{\mathcal I_j,\mathcal I_j}+\mathbf Z_i^{(j)},
\\
\bm\eta_{ij} &= \bm\kappa_i^{(j)} +\bm\Omega_{\mathcal I_j,\mathcal I_j}\bm\mu_{V,\mathcal I_j} -\bm\Omega_{\mathcal I_j,\mathcal I_{-j}}
\left(\mathbf V_{i,\mathcal I_{-j}}-\bm\mu_{V,\mathcal I_{-j}}\right).
\end{align*}

\subsection{Graph-Layer Updates}
\label{sec:graph_layer_updates}

\paragraph{Precision matrix.}
The conditional prior $p(\bm\mu_V\mid\bm\Omega)$ contributes an additional quadratic term and a factor $|\bm\Omega|^{1/2}$ to the precision-matrix update. Define
\[
    \mathbf S_{\Omega}^{*}
    =
    \sum_{i=1}^{N}
    \left(\mathbf V_i-\bm\mu_V\right)
    \left(\mathbf V_i-\bm\mu_V\right)^\top
    +
    \kappa_V
    \left(\bm\mu_V-\bm m_V\right)
    \left(\bm\mu_V-\bm m_V\right)^\top,
    \qquad
    n_{\Omega}=N+1.
\]
Conditional on $(\bm\alpha,\mathbf H)$, the off-diagonal entries of $\bm\Omega$ have Gaussian prior means and variances. For a within-node off-diagonal entry, the corresponding prior mean and variance are $0$ and $\sigma_{\mathrm{within}}^2$; for an entry in the block joining nodes $a$ and $b$, they are $\alpha_{ab}H_{ab,uv}$ and $\alpha_{ab}^2$, respectively.

Following \citet{Wang2015SSSL}, we permute a target column of $\bm\Omega$ to the last position and write
\[
    \bm\Omega
    =
    \begin{pmatrix}
        \bm\Omega_{11}&\mathbf u\\
        \mathbf u^\top&\omega_{22}
    \end{pmatrix},
    \qquad
    \mathbf S_{\Omega}^{*}
    =
    \begin{pmatrix}
        \mathbf S_{11}&\mathbf s_{12}\\
        \mathbf s_{12}^\top&s_{22}
    \end{pmatrix}.
\]
Let $\mathbf m_{12}$ collect the corresponding prior means, and let $\mathbf D_{12}$ be the diagonal matrix of the corresponding prior variances. Define
\[
    \gamma_{\Omega}
    =
    \omega_{22}
    -\mathbf u^\top\bm\Omega_{11}^{-1}\mathbf u,
    \qquad
    \mathbf C
    =
    \left\{
        (s_{22}+\lambda_{\mathrm{diag}})
        \bm\Omega_{11}^{-1}
        +\mathbf D_{12}^{-1}
    \right\}^{-1}.
\]
The column update is
\begin{align*}
    \mathbf u\mid\cdot
    &\sim
    \mathcal N_{Q-1}
    \left(
        -\mathbf C
        \left(
            \mathbf s_{12}
            -\mathbf D_{12}^{-1}\mathbf m_{12}
        \right),
        \mathbf C
    \right),\\
    \gamma_{\Omega}\mid\cdot
    &\sim
    \operatorname{Gamma}
    \left(
        \frac{n_{\Omega}}{2}+1,
        \frac{s_{22}+\lambda_{\mathrm{diag}}}{2}
    \right).
\end{align*}

\paragraph{Block-prior and graph-mean updates.}
Conditional on $\bm\Omega$ and $\bm\alpha$, the signed variables $H_{ab,uv}$ are sampled from their binary full conditional distributions. For each node pair, we update the shared multiplier using a Gaussian random-walk Metropolis--Hastings proposal, conditional on the current precision block, signed variables, and edge indicator.

When updating $G_{ab}$, we integrate out $\pi_{\mathrm{edge}}$ under its $\operatorname{Beta}(a_\pi,b_\pi)$ prior. Let $M_E=p(p-1)/2$, and let $E_{-ab}$ denote the number of included edges excluding $(a,b)$. The collapsed conditional log odds are
\begin{align*}
    \log
    \frac{\Pr(G_{ab}=1\mid\cdot)}
         {\Pr(G_{ab}=0\mid\cdot)}
    ={}&
    \log
    \frac{a_\pi+E_{-ab}}
         {b_\pi+M_E-1-E_{-ab}}
    +
    \log\left(\frac{\tau_0}{\tau_1}\right)
    +
    \frac{1}{2}
    \left(
        \frac{1}{\tau_0^2}
        -\frac{1}{\tau_1^2}
    \right)
    \alpha_{ab}^2.
\end{align*}

\section{Simulation Studies}
\label{sec:simulations}

We evaluated node-level graph recovery using data generated from Student-$t$ and Gaussian mixtures. We compared DGM with three benchmark methods under both data-generating mechanisms, fitting DGM with Student-\(t\) kernels throughout. We did not compare our method to M-LDQ because it is only applicable to univariate distributions and is a fundamentally different parametric model.

\subsection{Simulation Design}

We considered six settings formed by \(p\in\{20,40\}\) nodes and \(K_{\mathrm{true}}\in\{3,4,5\}\) mixture components. For each setting, we generated 50 data sets with \(N=200\) subjects, \(N_i^{(j)}=200\) observations per subject and node, and observation dimension \(d=10\) for each node. The graphs contained 21 and 41 edges for \(p=20\) and \(p=40\), respectively. The observations were generated from mixtures of Student-$t$ distributions with three degrees of freedom. For the sensitivity analysis, we generated data from Gaussian mixtures in the same settings to assess DGM graph recovery under mixture-kernel misspecification. Appendix~\ref{app:simulation-design} provides full details of data generation and the complete sensitivity-analysis results.

\subsection{Compared Methods}
\label{sec:sim_methods}

We compare the proposed DGM with three benchmark methods.

\paragraph{DGM.}
We set \(\nu_t=3\) for all DGM fits. For each simulation replicate, we fit DGM separately with \(K_{\mathrm{fit}}\in\{3,4,5\}\) components per node and select the value with the smallest widely applicable information criterion (WAIC). Each MCMC chain is run for \(3000\) iterations, with the first \(1500\) discarded as burn-in. We obtain a binary graph estimate by thresholding the posterior edge inclusion probabilities at \(0.5\).

\paragraph{Mean-glasso.}
For each subject, we compute the sample mean vector at each node and concatenate these vectors across nodes. We then apply the graphical lasso to the resulting features \citep{FriedmanHastieTibshirani2008}.

\paragraph{Mean-\(t\)-lasso.}
Using the same mean-vector representation as Mean-glasso, we fit a Student-\(t\) graphical lasso with three degrees of freedom \citep{FinegoldDrton2011}.

\paragraph{Multistage-MixT.}
At each node, a Student-\(t\) mixture is fitted independently using an expectation--maximization (EM) algorithm. The estimated subject-specific mixing weights are transformed to stick-breaking coordinates and used as inputs to a graphical lasso. The mixture size is fixed at the true value.

For the three benchmark methods, we standardized the resulting subject-level feature vectors across subjects before precision-matrix estimation. An edge at the node level was identified if the corresponding between-node block of the estimated precision matrix contained at least one nonzero entry.

We evaluated graph recovery at the node level using the Matthews correlation coefficient (MCC), F1 score, true positive rate (TPR), false positive rate (FPR), and structural Hamming distance (SHD).

\subsection{Simulation Results}
\label{sec:sim_results}

Tables~\ref{tab:sim_results_p20} and~\ref{tab:sim_results_p40} summarize the graph recovery for \(p=20\) and \(p=40\), respectively. Across all six settings, DGM achieves the highest mean MCC and F1 score and the lowest mean SHD. Its mean FPR does not exceed \(0.005\), while its mean TPR ranges from \(0.867\) to \(1\). Performance is similar for \(p=20\) and \(p=40\) when \(K_{\mathrm{true}}=3\) or \(4\). Recovery becomes more difficult when \(K_{\mathrm{true}}=5\), particularly for \(p=40\), where DGM has mean MCC \(0.910\) and mean F1 score \(0.912\).

\begin{table}[htbp]
\caption{Node-level graph recovery for \(p=20\) under the direct block partial-correlation design with 21 true edges, \(d_j=10\) for all nodes, \(N=200\) subjects, and \(N_i^{(j)}=200\) observations per subject--node pair. Entries are Monte Carlo means over 50 replicates, with standard deviations in parentheses.}
\label{tab:sim_results_p20}
\centering
\small
\setlength{\tabcolsep}{6pt}
\renewcommand{\arraystretch}{1.05}
\begin{tabular}{@{}c l ccccc@{}}
\toprule
\(K_{\mathrm{true}}\)
& Method & MCC & F1 & TPR & FPR & SHD \\
\midrule
\(3\)
& DGM
& 0.985 (0.019)
& 0.987 (0.017)
& 1.000 (0.000)
& 0.003 (0.004)
& 0.6 (0.8) \\
& Mean-glasso
& 0.066 (0.054)
& 0.210 (0.013)
& 0.969 (0.045)
& 0.907 (0.073)
& 153.9 (11.8) \\
& Mean-\(t\)-lasso
& 0.011 (0.037)
& 0.200 (0.002)
& 0.999 (0.007)
& 0.994 (0.010)
& 168.1 (1.7) \\
& Multistage-MixT
& 0.451 (0.121)
& 0.460 (0.114)
& 0.990 (0.026)
& 0.320 (0.137)
& 54.3 (23.2) \\
\addlinespace[4pt]
\(4\)
& DGM
& 0.973 (0.032)
& 0.976 (0.028)
& 0.979 (0.034)
& 0.003 (0.005)
& 1.0 (1.2) \\
& Mean-glasso
& 0.016 (0.064)
& 0.200 (0.008)
& 0.953 (0.045)
& 0.939 (0.029)
& 159.7 (4.6) \\
& Mean-\(t\)-lasso
& -0.005 (0.057)
& 0.199 (0.003)
& 0.995 (0.017)
& 0.997 (0.006)
& 168.6 (0.9) \\
& Multistage-MixT
& 0.377 (0.146)
& 0.327 (0.204)
& 0.315 (0.296)
& 0.040 (0.095)
& 21.2 (12.1) \\
\addlinespace[4pt]
\(5\)
& DGM
& 0.925 (0.054)
& 0.931 (0.050)
& 0.909 (0.071)
& 0.005 (0.006)
& 2.8 (2.0) \\
& Mean-glasso
& 0.030 (0.064)
& 0.202 (0.011)
& 0.940 (0.053)
& 0.913 (0.022)
& 155.6 (4.0) \\
& Mean-\(t\)-lasso
& 0.001 (0.006)
& 0.199 (0.000)
& 1.000 (0.000)
& 1.000 (0.002)
& 168.9 (0.3) \\
& Multistage-MixT
& 0.320 (0.119)
& 0.223 (0.138)
& 0.135 (0.099)
& 0.001 (0.004)
& 18.3 (1.8) \\
\bottomrule
\end{tabular}
\end{table}

\begin{table}[htbp]
\caption{Node-level graph recovery for \(p=40\) under the direct block partial-correlation design with 41 true edges, \(d_j=10\) for all nodes, \(N=200\) subjects, and \(N_i^{(j)}=200\) observations per subject--node pair. Entries are Monte Carlo means over 50 replicates, with standard deviations in parentheses.}
\label{tab:sim_results_p40}
\centering
\small
\setlength{\tabcolsep}{6pt}
\renewcommand{\arraystretch}{1.05}
\begin{tabular}{@{}c l ccccc@{}}
\toprule
\(K_{\mathrm{true}}\)
& Method & MCC & F1 & TPR & FPR & SHD \\
\midrule
\(3\)
& DGM
& 0.989 (0.011)
& 0.990 (0.011)
& 0.995 (0.010)
& 0.001 (0.001)
& 0.8 (0.9) \\
& Mean-glasso
& 0.055 (0.030)
& 0.109 (0.007)
& 0.939 (0.058)
& 0.851 (0.091)
& 631.5 (65.4) \\
& Mean-\(t\)-lasso
& 0.016 (0.029)
& 0.101 (0.003)
& 0.979 (0.029)
& 0.961 (0.038)
& 711.4 (27.3) \\
& Multistage-MixT
& 0.428 (0.130)
& 0.385 (0.135)
& 0.942 (0.121)
& 0.199 (0.114)
& 149.2 (84.0) \\
\addlinespace[4pt]
\(4\)
& DGM
& 0.981 (0.015)
& 0.981 (0.014)
& 0.980 (0.022)
& 0.001 (0.001)
& 1.5 (1.2) \\
& Mean-glasso
& 0.034 (0.030)
& 0.104 (0.004)
& 0.969 (0.027)
& 0.930 (0.027)
& 688.2 (19.7) \\
& Mean-\(t\)-lasso
& 0.007 (0.018)
& 0.100 (0.001)
& 0.998 (0.007)
& 0.995 (0.008)
& 735.1 (5.7) \\
& Multistage-MixT
& 0.314 (0.102)
& 0.260 (0.126)
& 0.215 (0.166)
& 0.015 (0.024)
& 43.2 (12.4) \\
\addlinespace[4pt]
\(5\)
& DGM
& 0.910 (0.044)
& 0.912 (0.043)
& 0.867 (0.066)
& 0.002 (0.002)
& 6.8 (3.2) \\
& Mean-glasso
& 0.038 (0.021)
& 0.104 (0.003)
& 0.963 (0.023)
& 0.917 (0.010)
& 679.5 (7.7) \\
& Mean-\(t\)-lasso
& 0.003 (0.016)
& 0.100 (0.000)
& 1.000 (0.003)
& 0.999 (0.002)
& 738.0 (1.3) \\
& Multistage-MixT
& 0.245 (0.079)
& 0.132 (0.080)
& 0.074 (0.052)
& 0.000 (0.001)
& 38.2 (1.9) \\
\bottomrule
\end{tabular}
\end{table}

Both mean-based methods select very dense graphs. Although their mean TPRs range from \(0.939\) to \(1.000\), their mean FPRs range from \(0.851\) to \(1.000\), resulting in MCC values close to zero and large SHD. Their high TPRs therefore primarily reflect the selection of most candidate edges rather than accurate graph recovery. Mean-\(t\)-lasso does not appreciably improve on Mean-glasso.

Multistage-MixT improves substantially on the mean-based methods in terms of MCC, F1, FPR, and SHD, but has lower mean MCC and F1 scores and larger mean SHD than DGM in every setting. When \(K_{\mathrm{true}}=3\), it retains high TPR at the cost of appreciable FPR. For \(K_{\mathrm{true}}=4\) and \(5\), its low FPR is accompanied by a marked reduction in TPR. Its MCC and F1 scores also exhibit greater Monte Carlo variability than those of DGM. This contrast is consistent with a benefit from jointly estimating the mixture representation and graph rather than treating estimated mixture weights as fixed inputs to a separate graph-estimation stage.

Appendix~\ref{app:simulation-gaussian-results} reports the complete results from the comparison using data generated from Gaussian mixtures. Repeating the comparison with data generated from Gaussian mixtures produces the same qualitative pattern. DGM again achieves the highest mean MCC and F1 score and the lowest mean SHD in all six settings. Its mean MCC ranges from \(0.901\) to \(0.991\). The largest decrease in mean MCC relative to the Student-\(t_3\) mixture results is \(0.009\), occurring at \(p=40\) and \(K_{\mathrm{true}}=5\). These results indicate that DGM retains accurate graph recovery under misspecification of the mixture-component distribution.

\section{Application to Single-Cell Data}
\label{sec:application}

We applied the proposed DGM to population-scale single-cell transcriptomic data from the OneK1K peripheral blood mononuclear cell cohort \citep{YazarEtAl2022}. The analysis included five major immune cell-type nodes: B cells, CD4 T cells, CD8 T cells, natural killer (NK) cells, and monocytes. The inferential target was an undirected graph of conditional dependence among donor-specific cell-type expression distributions. By restricting our analysis to ligand and receptor genes, the inferred conditional dependence can be interpreted as evidence of cell-cell communication across cell types.

We extract ligand and receptor genes from the CellChat database \citep{JinEtAl2021CellChat}. Each node therefore corresponded to a cell type, with each donor contributing a sample of multivariate expression profiles of those extracted ligand and receptor genes at that node.

We retained donors with at least 50 cells in each of the five cell types, yielding \(N=223\) donors. The node-specific feature dimensions (i.e., number of ligand and receptor genes) for B cells, CD4 T cells, CD8 T cells, NK cells, and monocytes were 27, 17, 20, 22, and 22, respectively.

\subsection{Mixture-Component Selection and Final Model Fitting}
\label{sec:application_fit}

For each node \(j\), we considered \(K_j\in\{3,4,5\}\) and fitted all \(3^5=243\) candidate models corresponding to the possible combinations of node-specific component counts, running each candidate model for 6000 MCMC iterations and discarding the first 3000 iterations as burn-in. We selected the combination of node-specific component counts with the smallest WAIC.

The selected combination of component counts was \((5,5,4,5,4)\) for B cells, CD4 T cells, CD8 T cells, NK cells, and monocytes, respectively. We subsequently refitted this model for 10,000 MCMC iterations, discarding the first 5,000 iterations and retaining 5,000 posterior draws for inference. The post-burn-in trace of the observed-data log-likelihood was stable, with rapidly decaying autocorrelation (Figure~\ref{fig:real-loglik-diagnostics} in Appendix~\ref{app:application-figures}).

\subsection{Posterior Graph}
\label{sec:application_graph}

Posterior support for each potential edge \((a,b)\) was summarized by the posterior inclusion probability \(\Pr(G_{ab}=1\mid\mathcal X)\). We ordered the PIPs from largest to smallest and selected the largest set of edges whose mean posterior noninclusion probability did not exceed \(q=0.01\). Appendix~\ref{sec:posterior-edge-selection} gives the formal selection rule. This procedure selected six edges, corresponding to an achieved Bayesian FDR of \(0.008\).

For posterior draw \(t\), let
\[
I_{ab\mid R_{ab}}^{(t)}=
-\frac{1}{2}
\sum_{\ell=1}^{m_{ab}}
\log\left\{
1-\left(\rho_{ab,\ell}^{(t)}\right)^2
\right\}
\]
denote the Gaussian conditional mutual information between latent blocks \(a\) and \(b\) given the remaining blocks \(R_{ab}\), where \(m_{ab}=\min\{K_a-1,K_b-1\}\) and \(\rho_{ab,\ell}^{(t)}\) is the \(\ell\)th block partial canonical correlation between latent blocks \(a\) and \(b\) given the remaining blocks \citep{ScharfWang2023}. To account for differences in the numbers of mixture components across nodes, we defined the dimension-normalized partial association measure
\[
\mathcal A_{ab\mid R_{ab}}^{(t)}
=
1-\exp\left\{
-\frac{2I_{ab\mid R_{ab}}^{(t)}}{m_{ab}}
\right\}
=
1-
\left[
\prod_{\ell=1}^{m_{ab}}
\left\{
1-\left(\rho_{ab,\ell}^{(t)}\right)^2
\right\}
\right]^{1/m_{ab}}.
\]
This measure adapts the geometric-mean multivariate association measure of \citet{CramerNicewander1979} to block partial canonical correlations. We report posterior medians and 95\% credible intervals of \(\mathcal A_{ab\mid R_{ab}}^{(t)}\) over all post-burn-in draws.

\begin{table}[htbp]
\centering
\caption{Posterior edge inclusion probabilities and dimension-normalized conditional-dependence measures under the selected DGM. The six edges above the horizontal rule were selected by controlling the posterior expected false-discovery proportion at \(q=0.01\). The dimension-normalized partial association measure \(\mathcal A_{ab\mid R_{ab}}\) is reported as the posterior median and 95\% credible interval. Values below \(0.001\) are reported as \(<0.001\).}
\label{tab:real_edge_summary}
\begingroup
\small
\setlength{\tabcolsep}{8pt}
\renewcommand{\arraystretch}{1.05}
\begin{tabular}{@{}lcc@{}}
\toprule
Cell-type pair
& PIP
& Median \(\mathcal A_{ab\mid R_{ab}}\) (95\% CrI) \\
\midrule
B cells -- CD4 T cells & 1.000 & 0.179 \((0.125,\ 0.240)\) \\
CD4 T cells -- CD8 T cells & 1.000 & 0.114 \((0.067,\ 0.173)\) \\
CD4 T cells -- monocytes & 1.000 & 0.048 \((0.018,\ 0.090)\) \\
CD8 T cells -- NK cells & 1.000 & 0.119 \((0.057,\ 0.191)\) \\
CD4 T cells -- NK cells & 0.977 & 0.036 \((<0.001,\ 0.072)\) \\
CD8 T cells -- monocytes & 0.973 & 0.036 \((<0.001,\ 0.092)\) \\
\midrule
NK cells -- monocytes & 0.909 & 0.023 \((<0.001,\ 0.076)\) \\
B cells -- NK cells & 0.859 & 0.024 \((<0.001,\ 0.089)\) \\
B cells -- CD8 T cells & 0.830 & 0.007 \((<0.001,\ 0.055)\) \\
B cells -- monocytes & 0.701 & 0.001 \((<0.001,\ 0.032)\) \\
\bottomrule
\end{tabular}
\endgroup
\end{table}

CD4 T cells had the highest degree in the selected graph, with edges to all four other cell types. The graph also contained CD8 T-cell--NK-cell and CD8 T-cell--monocyte edges, whereas B cells were adjacent only to CD4 T cells.

Dependence magnitudes differed substantially among edges with similar posterior inclusion probabilities. The B-cell--CD4 T-cell edge had the largest posterior median \(\mathcal A_{ab\mid R_{ab}}\) at 0.179, followed by the CD8 T-cell--NK-cell edge at 0.119 and the CD4 T-cell--CD8 T-cell edge at 0.114. Median values for the remaining selected edges ranged from 0.036 to 0.048. Thus, PIP and \(\mathcal A_{ab\mid R_{ab}}\) summarized posterior evidence for edge inclusion and the magnitude of conditional dependence, respectively.

\subsection{Biological Interpretation}
\label{sec:application_interpretation}

The B-cell--CD4 T-cell edge has a clear biological counterpart in the helper function of CD4 T cells during B-cell responses. T follicular helper cells, in particular, provide cognate signals that support germinal-center B-cell differentiation \citep{Crotty2011Tfh,MaEtAl2012Tfh}. The CD4 T-cell--CD8 T-cell edge similarly accords with the contribution of CD4 T-cell help to the development and maintenance of antiviral CD8 T-cell responses \citep{SwainEtAl2012CD4}.

The two NK-cell edges correspond to reported interactions with both T-cell compartments. NK cells regulate diverse CD4 and CD8 T-cell responses \citep{CromeEtAl2013NK}, while CD8 T cells and NK cells have complementary cytotoxic functions and documented functional interplay \citep{RosenbergHuang2018CD8NK,UzhachenkoShanker2019CTLNK}.

For the monocyte connections, circulating monocyte complexes containing either CD4 or CD8 T cells have been detected in human peripheral blood \citep{BurelEtAl2019TMonocyte}. Primary-cell studies further demonstrate nonadditive cytokine responses in CD4 T-cell--monocyte cocultures and enhanced CD8 T-cell stimulation by CD169\(^{+}\)CD14\(^{+}\) monocytes \citep{SchrierEtAl2016CD4Monocyte,AffandiEtAl2021CD169}.

\section{Discussion}
\label{sec:discussion}

We have developed a Bayesian framework for graph estimation with multivariate distribution-valued nodes observed through finite samples. The key idea of DGM is to represent the node-specific distributions using shared Student-\(t\) mixture components and subject-specific mixing weights, and to link transformed weights through a latent Gaussian graphical model. The block spike-and-slab construction enables edge selection at the node level, while the joint hierarchy accounts for uncertainty in the distributional representations during graph estimation.

Several limitations suggest directions for future research. First, the shared-component assumption requires all subjects at a given node to share a common set of component locations and covariances, with distributional heterogeneity represented through the mixing weights. Although this common dictionary facilitates cross-subject comparison and information sharing, it may be restrictive when component structure varies substantially across subjects. Second, the present framework estimates an undirected graph. Extending the latent graphical model to a directed acyclic graph would allow directed dependence structures among distribution-valued nodes to be investigated.

Although our application focused on single-cell RNA sequencing data, the framework applies to any setting with multiple distribution-valued variables in which characterizing their conditional dependence is of scientific interest.

\section*{Acknowledgements}
During the preparation of this manuscript, the authors utilized AI-assisted technologies. GPT-5.6 was employed for language polishing and editing. OpenAI Codex, using GPT-5.5, was used to assist in debugging and correcting the code. All AI-assisted outputs were reviewed, edited, and verified by the authors, who take full responsibility for the accuracy and integrity of the work.

\bibliographystyle{chicago}
\bibliography{reference}

\clearpage
\appendix
\renewcommand{\thesection}{\Alph{section}}
\renewcommand{\thesubsection}{\thesection.\arabic{subsection}}
\renewcommand{\theequation}{\thesection.\arabic{equation}}
\renewcommand{\thetable}{S\arabic{table}}
\renewcommand{\thefigure}{S\arabic{figure}}
\renewcommand{\theHsection}{appendix.\Alph{section}}
\renewcommand{\theHequation}{appendix.\thesection.\arabic{equation}}
\renewcommand{\theHtable}{appendix.S\arabic{table}}
\renewcommand{\theHfigure}{appendix.S\arabic{figure}}
\setcounter{table}{0}
\setcounter{figure}{0}
\numberwithin{equation}{section}

\section{Proof and Posterior Computation}
\label{app:posterior-computation}

This appendix gives a proof of Proposition~\ref{prop:mixture_identifiability} and the posterior-computation details for the covariance-matched Student-\(t\) distributional graphical model described in Sections~\ref{sec:model}--\ref{sec:inference}.

\subsection{Proof of Proposition~\ref{prop:mixture_identifiability}}
\label{app:proof-proposition-1}

\begin{proof}[Proof of Proposition~\ref{prop:mixture_identifiability}]
Finite mixtures of multivariate Student-\(t\) location--scatter densities are identifiable \citep{HolzmannMunkGneiting2006}. For fixed \(\nu_t>2\), the mapping \(\bm\Sigma\mapsto c_{\nu_t}\bm\Sigma\) is one-to-one, so the covariance-matched parameterization used in Section~\ref{sec:mixture} preserves identifiability.

Applying finite-mixture identifiability to \(f_1^{(j)}=\widetilde f_1^{(j)}\) yields a permutation \(\sigma_j\) that aligns the component parameter pairs and the mixing weights for subject \(1\). For any other subject \(i\), identifiability applied to \(f_i^{(j)}=\widetilde f_i^{(j)}\) yields a permutation aligning the same two shared dictionaries. Because the component parameter pairs are distinct, this alignment is unique and therefore must equal \(\sigma_j\). Hence,
\[
    \widetilde\pi_{i,\sigma_j(k)}^{(j)}
    =
    \pi_{i,k}^{(j)},
    \qquad
    i=1,\ldots,N,
    \quad
    k=1,\ldots,K_j.
\]
Thus, the component permutation is common to all subjects at node \(j\), which proves the proposition.
\end{proof}

\subsection{\texorpdfstring{Student-\(t\)}{Student-t} Mixture Layer}
\label{app:mixture-layer}

\subsubsection{Order-preserving component-parameter update}
\label{app:component-update}

For component $k$ at node $j$, define the current allocation set
\[
    \mathcal O_{jk}
    =
    \left\{
        (i,c):
        i=1,\ldots,N,\;
        c=1,\ldots,N_i^{(j)},\;
        z_{i,c}^{(j)}=k
    \right\},
    \qquad
    n_{jk}=|\mathcal O_{jk}|,
\]
and the precision weights
\[
    w_{i,c}^{(j)}=\frac{\xi_{i,c}^{(j)}}{c_{\nu_t}},
    \qquad
    W_{jk}=\sum_{(i,c)\in\mathcal O_{jk}}w_{i,c}^{(j)}.
\]
When $W_{jk}>0$, let
\begin{align*}
    \bar{\mathbf x}_{jk}^{w}
    &=
    \frac{1}{W_{jk}}
    \sum_{(i,c)\in\mathcal O_{jk}}
    w_{i,c}^{(j)}\mathbf x_{i,c}^{(j)},\\
    \mathbf S_{jk}^{w}
    &=
    \sum_{(i,c)\in\mathcal O_{jk}}
    w_{i,c}^{(j)}
    \left(\mathbf x_{i,c}^{(j)}-\bar{\mathbf x}_{jk}^{w}\right)
    \left(\mathbf x_{i,c}^{(j)}-\bar{\mathbf x}_{jk}^{w}\right)^\top.
\end{align*}
For an empty component, set $\bar{\mathbf x}_{jk}^{w}=\bm m_{0j}$ and $\mathbf S_{jk}^{w}=\mathbf 0$.

Under the Normal--Inverse-Wishart prior in Section~\ref{sec:priors}, define
\begin{align*}
    \kappa_{jk}^{*}
    &=\kappa_j+W_{jk},
    &
    \nu_{jk}^{*}
    &=\nu_j+n_{jk},\\
    \bm m_{jk}^{*}
    &=
    \frac{\kappa_j\bm m_{0j}
    +W_{jk}\bar{\mathbf x}_{jk}^{w}}
    {\kappa_j+W_{jk}},
\end{align*}
and
\[
    \bm\Psi_{jk}^{*}
    =
    \bm\Psi_j+\mathbf S_{jk}^{w}+
    \frac{\kappa_jW_{jk}}{\kappa_j+W_{jk}}
    \left(\bar{\mathbf x}_{jk}^{w}-\bm m_{0j}\right)
    \left(\bar{\mathbf x}_{jk}^{w}-\bm m_{0j}\right)^\top.
\]
Given the current component mean $\bm\mu_{j,k}^{\mathrm{cur}}$, draw
\[
    \bm\Sigma_{j,k}\mid\bm\mu_{j,k}^{\mathrm{cur}},\cdot
    \sim
    \mathcal{IW}_{d_j}
    \left(
        \nu_{jk}^{*}+1,
        \bm\Psi_{jk}^{*}
        +\kappa_{jk}^{*}
        \left(\bm\mu_{j,k}^{\mathrm{cur}}-\bm m_{jk}^{*}\right)
        \left(\bm\mu_{j,k}^{\mathrm{cur}}-\bm m_{jk}^{*}\right)^\top
    \right).
\]
Given the updated covariance matrix \(\bm\Sigma_{j,k}\), define
\[
    \mathbf B_{jk}
    =
    \frac{\bm\Sigma_{j,k}}{\kappa_{jk}^{*}}.
\]

For \(d_j>1\), partition
\[
    \bm\mu_{j,k}
    =
    \begin{pmatrix}
        \mu_{j,k,1}\\
        \bm\mu_{j,k,-1}
    \end{pmatrix},
    \qquad
    \bm m_{jk}^{*}
    =
    \begin{pmatrix}
        m_{jk,1}^{*}\\
        \bm m_{jk,-1}^{*}
    \end{pmatrix},
    \qquad
    \mathbf B_{jk}
    =
    \begin{pmatrix}
        B_{jk,11} & \mathbf B_{jk,1,-1}\\
        \mathbf B_{jk,-1,1} & \mathbf B_{jk,-1,-1}
    \end{pmatrix}.
\]

With the conventions \(\mu_{j,0,1}=-\infty\) and \(\mu_{j,K_j+1,1}=+\infty\), we first sample the first coordinate from its marginal full conditional,
\[
    \mu_{j,k,1}
    \mid
    \bm\Sigma_{j,k},\cdot
    \sim
    \operatorname{TN}
    \left(
        m_{jk,1}^{*},
        B_{jk,11};
        \mu_{j,k-1,1},
        \mu_{j,k+1,1}
    \right),
\]
where \(\operatorname{TN}(m,s^2;L,U)\) denotes a univariate normal distribution with mean \(m\) and variance \(s^2\), truncated to \((L,U)\).

Conditional on the sampled first coordinate, the remaining coordinates are drawn from
\[
\begin{aligned}
    \bm\mu_{j,k,-1}
    \mid
    \mu_{j,k,1},\bm\Sigma_{j,k},\cdot
    \sim
    \mathcal N_{d_j-1}
    \Bigg(
    &\bm m_{jk,-1}^{*}
    +
    \mathbf B_{jk,-1,1}B_{jk,11}^{-1}
    \left(
        \mu_{j,k,1}-m_{jk,1}^{*}
    \right),\\
    &\mathbf B_{jk,-1,-1}
    -
    \mathbf B_{jk,-1,1}B_{jk,11}^{-1}
    \mathbf B_{jk,1,-1}
    \Bigg).
\end{aligned}
\]

For the numerical studies, \(\bm m_{0j}\) is the pooled sample mean at node \(j\), \(\kappa_j=0.01\), \(\nu_j=d_j+2\), and \(\bm\Psi_j\) is the pooled sample covariance.

\subsection{Graph Mean and Precision Matrix}
\label{app:graph-update}

\subsubsection{Latent graph mean}

Under the prior specified in Section~\ref{sec:priors}, the graph mean is sampled from
\[
    \bm\mu_V\mid\{\mathbf V_i\}_{i=1}^N,\bm\Omega
    \sim
    \mathcal N_Q\left(
        \frac{\kappa_V\bm m_V+N\bar{\mathbf V}}
             {\kappa_V+N},
        \{(\kappa_V+N)\bm\Omega\}^{-1}
    \right),
\]
where \(\bar{\mathbf V}=N^{-1}\sum_{i=1}^N\mathbf V_i\).

\subsubsection{Derivation of the column-wise precision update}

Using the column partition and prior moments defined in Section~\ref{sec:graph_layer_updates} and the column-wise construction of \citet{Wang2015SSSL}, write
\[
    \gamma_{\Omega}
    =
    \omega_{22}
    -
    \mathbf u^\top
    \bm\Omega_{11}^{-1}
    \mathbf u.
\]
The Schur-complement identities give
\[
    |\bm\Omega|
    =
    |\bm\Omega_{11}|\gamma_{\Omega},
    \qquad
    \omega_{22}
    =
    \gamma_{\Omega}
    +
    \mathbf u^\top
    \bm\Omega_{11}^{-1}
    \mathbf u,
\]
and the transformation from \((\mathbf u,\omega_{22})\) to \((\mathbf u,\gamma_{\Omega})\) has unit Jacobian. Conditional on the remaining quantities, the resulting kernel is
\[
\begin{aligned}
p(\mathbf u,\gamma_{\Omega}\mid\cdot)
&\propto
\gamma_{\Omega}^{\,n_{\Omega}/2}
\exp\left\{
    -\frac{s_{22}+\lambda_{\mathrm{diag}}}{2}
    \gamma_{\Omega}
\right\}
\mathbf 1\{\gamma_{\Omega}>0\}
\\
&\quad\times
\exp\left[
    -\frac12
    \left\{
        \mathbf u^\top\mathbf C^{-1}\mathbf u
        +
        2\left(
            \mathbf s_{12}
            -
            \mathbf D_{12}^{-1}\mathbf m_{12}
        \right)^\top
        \mathbf u
    \right\}
\right],
\end{aligned}
\]
where
\[
    \mathbf C^{-1}
    =
    (s_{22}+\lambda_{\mathrm{diag}})
    \bm\Omega_{11}^{-1}
    +
    \mathbf D_{12}^{-1}.
\]
Thus the kernel factors into a Gaussian kernel in \(\mathbf u\) and a Gamma kernel in \(\gamma_{\Omega}\), yielding the full conditional distributions given in Section~\ref{sec:graph_layer_updates}.

\subsection{Signed Variables and Shared Block Multipliers}
\label{app:block-update}

\subsubsection{Signed variables}

Because the two signs have equal prior probability,
\[
    \logit\Pr(H_{ab,uv}=+1\mid\cdot)
    =
    \frac{
        (\omega_{ab,uv}+\alpha_{ab})^2
        -(\omega_{ab,uv}-\alpha_{ab})^2
    }{
        2\alpha_{ab}^{2}
    }.
\]

\subsubsection{Shared block multipliers}

Conditional on the current precision block, signed variables, and \(G_{ab}\),
\[
p(\alpha_{ab}\mid\cdot)
\propto
p(\alpha_{ab}\mid G_{ab})
\prod_{u=1}^{r_a}\prod_{v=1}^{r_b}
p(\omega_{ab,uv}\mid\alpha_{ab},H_{ab,uv}).
\]
Hence, up to an additive constant, the conditional log-kernel is
\[
\ell_{ab}(\alpha_{ab})
=
-\frac{\alpha_{ab}^2}{2\tau_{G_{ab}}^2}
-\frac{r_ar_b}{2}\log\alpha_{ab}^{2}
-\frac{1}{2\alpha_{ab}^{2}}
\sum_{u=1}^{r_a}\sum_{v=1}^{r_b}
\left(\omega_{ab,uv}-\alpha_{ab}H_{ab,uv}\right)^2.
\]
A Gaussian random walk on the log magnitude is given by
\[
\log|\alpha_{ab}^{\mathrm{prop}}|
=
\log|\alpha_{ab}|+e,
\qquad
e\sim\mathcal N(0,s_\alpha^2),
\]
with \(\operatorname{sign}(\alpha_{ab}^{\mathrm{prop}})=\operatorname{sign}(\alpha_{ab})\). The corresponding log Metropolis--Hastings ratio is
\[
\ell_{ab}(\alpha_{ab}^{\mathrm{prop}})
+\log|\alpha_{ab}^{\mathrm{prop}}|
-\ell_{ab}(\alpha_{ab})
-\log|\alpha_{ab}|,
\]
where the final two terms account for the Jacobian of the log-magnitude transformation.

\section{Simulation Design and Additional Results}
\label{app:simulation-design}

This appendix gives the data-generating mechanism, model-fitting details, and Gaussian-mixture sensitivity results for the simulation studies in Section~\ref{sec:simulations}.

\subsection{Block Precision Matrix}
\label{app:simulation-precision}

Let \(r_j=r=K_{\mathrm{true}}-1\) for all nodes and \(Q=pr\). For each replicate, a random graph with the prescribed number of edges was generated. The diagonal blocks of the latent precision matrix \(\bm\Omega_0\in\mathbb R^{Q\times Q}\) were set to \((\bm\Omega_0)_{\mathcal I_a,\mathcal I_a}=c_0\mathbf I_r\), \(a=1,\ldots,p\), with \(c_0=6\). For each true edge \((a,b)\) and \(u,v=1,\ldots,r\), an unnormalized block \(\mathbf U_{ab}^{(0)}\) was generated entrywise as
\[
\begin{aligned}
U_{ab,uv}^{(0)}&=H_{ab,uv}^{(0)}M_{ab,uv},\\
\Pr\{H_{ab,uv}^{(0)}=1\}&=\Pr\{H_{ab,uv}^{(0)}=-1\}=\tfrac12,\qquad M_{ab,uv}\sim\operatorname{Unif}(0.8,1.2).
\end{aligned}
\]
Setting \(\mathbf U_{ab}=\mathbf U_{ab}^{(0)}/\|\mathbf U_{ab}^{(0)}\|_F\) and \(s_{ab}\sim\operatorname{Unif}(0.40,0.60)\), the corresponding off-diagonal blocks were set to
\[
(\bm\Omega_0)_{\mathcal I_a,\mathcal I_b}=c_0s_{ab}\mathbf U_{ab},\qquad (\bm\Omega_0)_{\mathcal I_b,\mathcal I_a}=c_0s_{ab}\mathbf U_{ab}^{\top}.
\]
All nonedge blocks were zero. Consequently, every true edge satisfied
\[
\left\|\left\{(\bm\Omega_0)_{\mathcal I_a,\mathcal I_a}\right\}^{-1/2}(\bm\Omega_0)_{\mathcal I_a,\mathcal I_b}\left\{(\bm\Omega_0)_{\mathcal I_b,\mathcal I_b}\right\}^{-1/2}\right\|_F=s_{ab}.
\]
For a fixed collection of signal strengths \(\{s_{ab}\}\), block shapes were redrawn up to 10,000 times until \(\lambda_{\min}(\bm\Omega_0)>10^{-4}\). The signal strengths were redrawn after an unsuccessful set of block-shape proposals, with at most 50 such sets. This construction fixes the graph at the node level and directly controls the strength of each active between-node block.

\subsection{Latent Coordinates and Component Parameters}
\label{app:simulation-observations}

The subject-level latent coordinates were generated as
\[
\mathbf V_i\sim\mathcal N_Q(\bm\mu_{V,0},\bm\Omega_0^{-1}),\qquad i=1,\ldots,N,
\]
where each node-specific block of \(\bm\mu_{V,0}\) is the inverse logistic stick-breaking representation of uniform mixture weights. Applying the logistic stick-breaking map to each block of \(\mathbf V_i\) yielded the subject-specific mixing weights.

The component means were constructed to be ordered, with a gap of \(2\) between adjacent components. Component covariances followed a common AR(1) structure with marginal standard deviation \(0.6\) and correlation parameter \(0.1\).

\subsection{DGM Fitting Settings}
\label{app:simulation-fitting}

Table~\ref{tab:posterior-settings} summarizes the prior and posterior-computation settings used for each fixed-\(K\) DGM fit.

\begin{table}[htbp]
\centering
\caption{Posterior-computation settings for the DGM simulation fits.}
\label{tab:posterior-settings}
\begingroup
\small
\setlength{\tabcolsep}{8pt}
\renewcommand{\arraystretch}{1.05}
\begin{tabular}{@{}lll@{}}
\toprule
Quantity & Symbol & Value \\
\midrule
Student-$t$ degrees of freedom & $\nu_t$ & $3$ \\
Graph-mean prior mean & $\bm m_V$ & $\mathbf 0$ \\
Graph-mean prior precision multiplier & $\kappa_V$ & $1$ \\
Within-node off-diagonal variance & $\sigma_{\mathrm{within}}^2$ & $0.10$ \\
Diagonal exponential rate control & $\lambda_{\mathrm{diag}}$ & $1$ \\
Spike and slab scales & $(\tau_0,\tau_1)$ & $(0.05,1.00)$ \\
Sparsity hyperprior parameters & $(a_\pi,b_\pi)$ & $(1,1)$ \\
Log-multiplier proposal SD & $s_\alpha$ & $0.50$ \\
Posterior edge threshold & $c_{\mathrm{edge}}$ & $0.50$ \\
\bottomrule
\end{tabular}
\endgroup
\end{table}

\subsection{Gaussian-Mixture Sensitivity Results}
\label{app:simulation-gaussian-results}

Tables~\ref{tab:sim_results_gauss_p20} and~\ref{tab:sim_results_gauss_p40} report node-level graph recovery when the data were generated from Gaussian mixtures. All four methods were fitted using the same procedures as in the Student-$t_3$ simulations. Across the six Gaussian-mixture settings, DGM has the highest mean MCC and F1 score and the lowest mean SHD.

\begin{table}[htbp]
\caption{Node-level graph recovery for $p=20$ with Gaussian mixture data under the direct block partial-correlation design with 21 true edges, $d_j=10$ for all nodes, $N=200$ subjects, and $N_i^{(j)}=200$ observations per subject--node pair.}
\label{tab:sim_results_gauss_p20}
\centering
\small
\setlength{\tabcolsep}{6pt}
\renewcommand{\arraystretch}{1.05}
\begin{tabular}{@{}c l ccccc@{}}
\toprule
$K_{\mathrm{true}}$
& Method & MCC & F1 & TPR & FPR & SHD \\
\midrule
$3$
& DGM
& 0.988 (0.019)
& 0.989 (0.017)
& 0.998 (0.009)
& 0.002 (0.004)
& 0.5 (0.7) \\
& Mean-glasso
& 0.040 (0.079)
& 0.207 (0.014)
& 0.959 (0.046)
& 0.914 (0.073)
& 155.3 (12.1) \\
& Mean-$t$-lasso
& -0.009 (0.054)
& 0.198 (0.003)
& 0.994 (0.016)
& 0.997 (0.004)
& 168.6 (0.7) \\
& Multistage-MixT
& 0.448 (0.119)
& 0.455 (0.118)
& 0.970 (0.135)
& 0.309 (0.143)
& 52.8 (23.4) \\
\addlinespace[4pt]
$4$
& DGM
& 0.977 (0.025)
& 0.979 (0.023)
& 0.983 (0.030)
& 0.003 (0.005)
& 0.9 (1.0) \\
& Mean-glasso
& 0.040 (0.056)
& 0.203 (0.008)
& 0.969 (0.045)
& 0.941 (0.023)
& 159.6 (3.7) \\
& Mean-$t$-lasso
& 0.000 (0.044)
& 0.199 (0.002)
& 0.996 (0.013)
& 0.996 (0.005)
& 168.5 (0.9) \\
& Multistage-MixT
& 0.396 (0.135)
& 0.344 (0.193)
& 0.332 (0.295)
& 0.043 (0.099)
& 21.3 (12.6) \\
\addlinespace[4pt]
$5$
& DGM
& 0.928 (0.045)
& 0.934 (0.041)
& 0.912 (0.063)
& 0.005 (0.006)
& 2.7 (1.6) \\
& Mean-glasso
& 0.049 (0.060)
& 0.206 (0.010)
& 0.956 (0.044)
& 0.912 (0.020)
& 155.0 (3.9) \\
& Mean-$t$-lasso
& 0.000 (0.000)
& 0.199 (0.000)
& 1.000 (0.000)
& 1.000 (0.000)
& 169.0 (0.0) \\
& Multistage-MixT
& 0.332 (0.098)
& 0.235 (0.114)
& 0.141 (0.084)
& 0.001 (0.005)
& 18.3 (1.4) \\
\bottomrule
\end{tabular}
\end{table}

\begin{table}[htbp]
\caption{Node-level graph recovery for $p=40$ with Gaussian mixture data under the direct block partial-correlation design with 41 true edges, $d_j=10$ for all nodes, $N=200$ subjects, and $N_i^{(j)}=200$ observations per subject--node pair.}
\label{tab:sim_results_gauss_p40}
\centering
\small
\setlength{\tabcolsep}{6pt}
\renewcommand{\arraystretch}{1.05}
\begin{tabular}{@{}c l ccccc@{}}
\toprule
$K_{\mathrm{true}}$
& Method & MCC & F1 & TPR & FPR & SHD \\
\midrule
$3$
& DGM
& 0.991 (0.012)
& 0.991 (0.011)
& 0.997 (0.008)
& 0.001 (0.001)
& 0.7 (0.9) \\
& Mean-glasso
& 0.057 (0.026)
& 0.110 (0.007)
& 0.936 (0.056)
& 0.844 (0.083)
& 626.6 (59.9) \\
& Mean-$t$-lasso
& 0.014 (0.033)
& 0.101 (0.003)
& 0.977 (0.027)
& 0.962 (0.038)
& 712.1 (27.6) \\
& Multistage-MixT
& 0.424 (0.127)
& 0.380 (0.135)
& 0.943 (0.133)
& 0.199 (0.112)
& 149.6 (81.8) \\
\addlinespace[4pt]
$4$
& DGM
& 0.979 (0.016)
& 0.980 (0.015)
& 0.977 (0.021)
& 0.001 (0.001)
& 1.6 (1.2) \\
& Mean-glasso
& 0.034 (0.028)
& 0.104 (0.004)
& 0.970 (0.030)
& 0.931 (0.025)
& 689.2 (18.2) \\
& Mean-$t$-lasso
& 0.005 (0.025)
& 0.100 (0.001)
& 0.998 (0.007)
& 0.995 (0.009)
& 735.2 (6.7) \\
& Multistage-MixT
& 0.290 (0.110)
& 0.235 (0.144)
& 0.207 (0.191)
& 0.016 (0.023)
& 44.0 (10.8) \\
\addlinespace[4pt]
$5$
& DGM
& 0.901 (0.040)
& 0.903 (0.040)
& 0.855 (0.068)
& 0.002 (0.002)
& 7.4 (2.8) \\
& Mean-glasso
& 0.036 (0.021)
& 0.104 (0.003)
& 0.961 (0.024)
& 0.918 (0.011)
& 679.9 (8.3) \\
& Mean-$t$-lasso
& 0.005 (0.005)
& 0.100 (0.000)
& 1.000 (0.000)
& 0.999 (0.001)
& 738.2 (1.0) \\
& Multistage-MixT
& 0.244 (0.094)
& 0.131 (0.080)
& 0.072 (0.047)
& 0.000 (0.000)
& 38.1 (1.9) \\
\bottomrule
\end{tabular}
\end{table}

\section{Application Details}
\label{app:application-details}

\subsection{Ligand--Receptor Feature Construction}
\label{app:lr-features}

The ligand--receptor interactions among the five cell types were obtained from CellChat \citep{JinEtAl2021CellChat,JinPlikusNie2025CellChat}. Let \(\mathcal L_{s\to t}\) and \(\mathcal R_{s\to t}\) denote, respectively, the ligand and receptor genes in an interaction from source cell type \(s\) to target cell type \(t\). The feature set for node \(j\) was
\begin{equation}
    \mathcal P_j^{\mathrm{LR}}
    =
    \bigcup_{t\ne j}\mathcal L_{j\to t}
    \ \cup\
    \bigcup_{s\ne j}\mathcal R_{s\to j}.
    \label{eq:lr-union}
\end{equation}
All genes in the resulting sets were present in the expression matrices. The final node-specific feature sets are listed in Table~\ref{tab:lr-genes}.

\begingroup
\small
\begin{longtable}{@{}p{1.15in}p{5.05in}@{}}
\caption{Ligand--receptor-informed feature sets for the five immune cell-type nodes.}
\label{tab:lr-genes}\\
\toprule
Node & Genes \\
\midrule
\endfirsthead
\multicolumn{2}{l}{\tablename~\thetable\ continued} \\
\toprule
Node & Genes \\
\midrule
\endhead
\bottomrule
\endfoot
B cells &
HLA-F, HLA-B, HLA-C, HLA-E, HLA-A, CD69, CLEC2D, PPIA, MIF,
CD99, HLA-DRA, HLA-DPB1, HLA-DPA1, HLA-DRB1, HLA-DQB1,
HLA-DMA, HLA-DMB, HLA-DQA1, HLA-DOB, CD74, CXCR4, CD44, CD22,
TNFRSF13C, TNFRSF14, PTPRC, SELL \\
\addlinespace
CD4 T cells &
TNFRSF14, HLA-B, HLA-C, HLA-E, MIF, HLA-A, LCK, CD69, CLEC2D,
PPIA, CD99, HLA-F, PTPRC, ANXA1, CD44, SELL, P4HB \\
\addlinespace
CD8 T cells &
BSG, HLA-E, MIF, CD99, CLEC2D, CLEC2B, CD69, PPIA, PTPRC,
ANXA1, HLA-DPB1, HLA-F, HLA-DPA1, CD8B, CD8A, KLRK1, CD74,
CXCR4, CD44, P4HB \\
\addlinespace
NK cells &
ADGRE5, HLA-B, HLA-C, HLA-E, HLA-A, MIF, CD99, PPIA, HLA-F,
LCK, PTPRC, CD160, ANXA1, HLA-DPB1, KLRK1, KLRD1, KLRC1,
KLRB1, CD74, CD44, BSG, P4HB \\
\addlinespace
Monocytes &
TNFRSF14, HLA-B, HLA-C, HLA-E, HLA-A, MIF, LGALS9, CLEC2B,
PPIA, CD99, HLA-F, PTPRC, SELPLG, TNFSF13B, CD55, CD74, CD44,
BSG, PILRA, CD4, FPR1, LILRB1 \\
\end{longtable}
\endgroup

\subsection{Posterior Edge Selection}
\label{sec:posterior-edge-selection}

For \(T\) retained posterior draws, the posterior inclusion probability for edge \((a,b)\) was estimated by
\begin{equation}
    \widehat p_{ab}
    =
    \frac{1}{T}
    \sum_{t=1}^{T}G_{ab}^{(t)}.
    \label{eq:application-pip}
\end{equation}
Let \(M_E=p(p-1)/2\) and order these probabilities as \(\widehat p_{(1)}\ge\cdots\ge\widehat p_{(M_E)}\). For a candidate set containing the first \(m\) edges, the posterior expected false-discovery proportion was estimated by
\begin{equation}
    \widehat{\operatorname{BFDR}}(m)
    =
    \frac{1}{m}
    \sum_{\ell=1}^{m}
    \left\{1-\widehat p_{(\ell)}\right\}.
    \label{eq:application-bfdr}
\end{equation}
The selected number of edges was
\begin{equation}
    \widehat m
    =
    \max\left\{
        m\in\{1,\ldots,M_E\}:
        \widehat{\operatorname{BFDR}}(m)\le q
    \right\}.
    \label{eq:application-bfdr-rule}
\end{equation}
We set \(q=0.01\) in the application.

\subsection{Application Diagnostics}
\label{app:application-figures}

\begin{figure}[htbp]
\centering
\textbf{(a)}\par\smallskip
\includegraphics[width=0.88\linewidth]{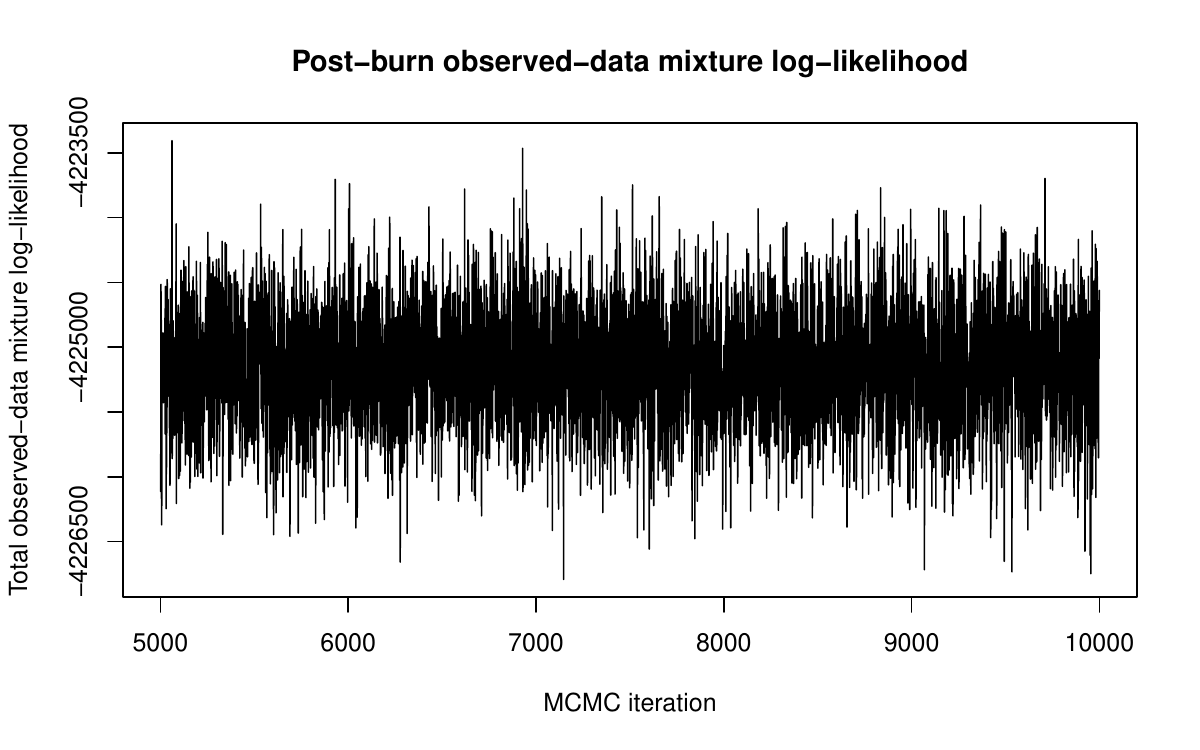}

\vspace{1em}
\textbf{(b)}\par\smallskip
\includegraphics[width=0.88\linewidth]{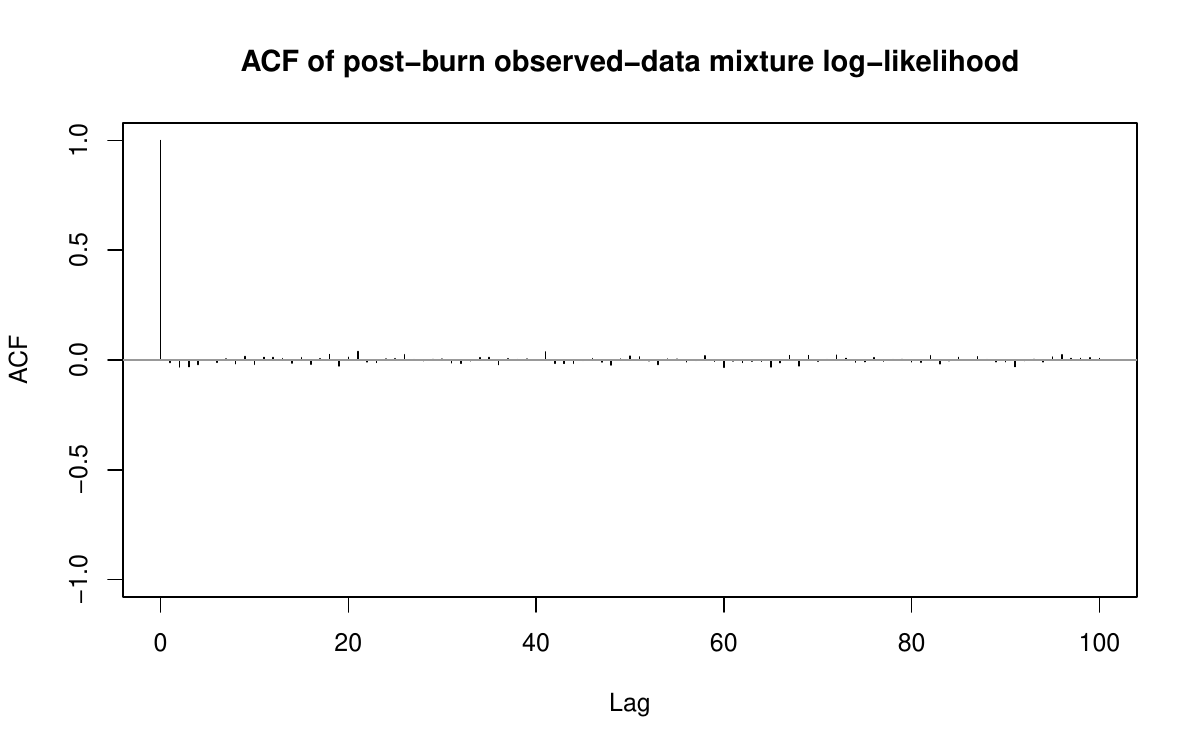}
\caption{Post-burn-in diagnostics for the observed-data log likelihood: (a) trace plot and (b) autocorrelation function.}
\label{fig:real-loglik-diagnostics}
\end{figure}

Figure~\ref{fig:real-loglik-diagnostics} gives the post-burn-in trace and autocorrelation function of the observed-data log likelihood.

\end{document}